\documentclass[aps,pre,twocolumn,superscriptaddress]{revtex4-2}

\usepackage{amsmath,amssymb}
\usepackage{times,hyperref,comment}
\usepackage{xcolor}
\hypersetup{
    colorlinks=true,       % false: boxed links; true: colored links
    linkcolor=blue,          % color of internal links (change box color with linkbordercolor)
    citecolor=blue,        % color of links to bibliography
    filecolor=blue,      % color of file links
    urlcolor=blue          % color of external links
}

\renewcommand{\d}{{\text{d}}}
\newcommand{\grad}{{\boldsymbol{\nabla}}}
\newcommand{\T}{\mathrm{T}}

\newcommand{\ie}{i.e., }
\newcommand{\eg}{e.g., }

\newcommand{\vecA}{\mathbf{A}}

\newcommand{\vecb}{\mathbf{b}}
\newcommand{\vecD}{\mathbf{D}}
\newcommand{\vecF}{\mathbf{F}}
\newcommand{\vech}{\mathbf{h}}
\newcommand{\vecI}{\mathbf{I}}
\newcommand{\vecJ}{\mathbf{J}}

\newcommand{\veco}{\mathbf{0}}

\newcommand{\vecX}{\mathbf{X}}
\newcommand{\vecx}{\mathbf{x}}
\newcommand{\vecy}{\mathbf{y}}

\newcommand{\calP}{\mathcal{P}}
\newcommand{\lag}{\hat{L}}

\newcommand{\pertH}{{\tilde{H}}}
\newcommand{\pertPhi}{{\tilde{\Phi}}}
\newcommand{\kt}{k_\mathrm{B}T}
\newcommand{\kb}{k_\mathrm{B}}
\newcommand{\teff}{\mathcal{T}}

\newcommand{\ccc}{{\mathrm{c}}}

\newcommand{\diss}{R}
\newcommand{\tf}{t_\mathrm{f}}

\newcommand{\noise}{\boldsymbol{\eta}}
\newcommand{\bmu}{\boldsymbol{\mu}}

\usepackage{xcolor}

\newcommand{\lefts}{\overleftarrow}
\newcommand{\rights}{\overrightarrow}

\begin{document} 

\title{Second Law of Thermodynamics without Einstein Relation} 

\author{Benjamin Sorkin} % 0000-0003-2799-3797

\thanks{Present address: Princeton Center for Theoretical Science, Princeton University, 08544 Princeton, New Jersey, USA.}
\affiliation{School of Chemistry and Center for Physics and Chemistry of Living Systems, Tel Aviv University, 69978 Tel Aviv, Israel}

\author{Haim Diamant} % 0000-0003-3858-1114

\affiliation{School of Chemistry and Center for Physics and Chemistry of Living Systems, Tel Aviv University, 69978 Tel Aviv, Israel}

\author{Gil Ariel} % 0000-0002-7251-5383

\affiliation{Department of Mathematics, Bar-Ilan University, 52000 Ramat Gan, Israel}

\author{Tomer Markovich}

\email{tmarkovich@tauex.tau.ac.il}

\affiliation{School of Mechanical Engineering and Center for Physics and Chemistry of Living Systems, Tel Aviv University, 69978 Tel Aviv, Israel}

\begin{abstract}

Materials that are constantly driven out of thermodynamic equilibrium, such as active and living systems, typically violate the Einstein relation. This may arise from active contributions to particle fluctuations which are unrelated to the dissipative resistance of the surrounding medium.
We show that in these cases the widely used relation between informatic entropy production and heat dissipation does not hold. 
Consequently, fluctuation relations for the mechanical work, such as the Jarzynski and Crooks theorems, are invalid. We relate the breaking of the correspondence between entropy production and heat dissipation to departure from the fluctuation-dissipation theorem.
We propose a temperaturelike variable that restores this correspondence and gives rise to a generalized second law of thermodynamics, whereby the dissipated heat is necessarily non-negative and vanishes at equilibrium.
The Clausius inequality, Carnot maximum efficiency theorem, and relation between the  extractable work and the change of free energy are recovered as well.

\end{abstract}

\maketitle

Classical thermodynamics is the central theory for the possible forms of energy transfer in materials. Its many useful consequences, including the Carnot efficiency theorem and the bounds on extractable work via changes in free energy, establish nontrivial and universal relations based on  a few macroscopic observables. Among these variables is the temperature, which ties  microscopic statistics of equilibrium systems to their energetics, e.g., through the equipartition principle or by imposing the mean energy as a constraint for entropy maximization.

In light of its vast applicability for characterizing the behaviors of equilibrium materials, there is an ongoing effort to extend  thermodynamics to systems that are constantly driven out of equilibrium, for example, self-propelled particles~\cite{Rev:Ram10,Rev:MarRam12,REV:bacteria}, sustained chemical reactions~\cite{SiefertJCP07,RaoJCP18}, and living systems~\cite{GovRev18}. Among the complications associated with nonequilibrium systems is the inadequacy of temperature and other variables as thermodynamic state functions~\cite{CasasVazquez2003,Cugliandolo2011,TakatoriPRE15,solon2015pressure,vanRoij2018,SpeckEL2016}. 

One of the consequences of thermodynamic equilibrium, where the system's configurations are Boltzmann distributed, is the Einstein relation (ER)~\cite{Einstein1905,KuboRPP66},
\begin{equation}
    \vecD= \kb T\bmu \, .\label{eq:einstein}
\end{equation}
The diffusivity $\vecD$ and mobility $\bmu$ characterize, respectively, the strength of thermal fluctuations and  dissipative attenuation of motion. In general they may be space-dependent matrices, reflecting translational and rotational symmetry breaking. At equilibrium  with a bath of temperature $T$, the effects of fluctuations and dissipation exactly balance~\cite{KuboRPP66}, leading to Eq.~\eqref{eq:einstein}. 
Systems away from equilibrium, such as supercooled liquids~\cite{ChenPNAS06} and systems under external forces~\cite{Cugliandolo2011}, typically do not satisfy Eq.~\eqref{eq:einstein} with the ambient temperature $T$. 
In such cases Eq.~\eqref{eq:einstein} has been used to define an effective temperature~\cite{CasasVazquez2003,solon2015pressure,CapriniCP24,DimaYael24,GalorYael24}. However, if the forces are not conservative~\cite{Cugliandolo2011,SuriPNAS18}, or when fluctuations arise from coarse graining of fast athermal or active forces~\cite{CuglianoloPRE97,CugliandoloJSTAT05,GalorYael24,DimaYael24,fodor2016EP,solon2015pressure,GranickSOFT20,PalacciPRL10,GranekPRL22,SolonJPA22}, there is generally no relation between the two position-dependent tensors $\vecD$ and $\bmu$. Thus, 
in far-from-equilibrium cases such as active biopolymer networks~\cite{BasuEPJE08,FredSCI07,Chen2020}, bacterial swarms~\cite{ChenPRL07,MaggiPRL14}, and driven chemical reactions~\cite{LacostePRL07,RaoJCP18}, the ER is invalid, as no scalar $T$ exists to satisfy Eq.~\eqref{eq:einstein}. In this Letter, we propose a temperaturelike variable for systems exhibiting this severe breakdown of the Einstein relation.

One of the theories addressing nonequilibrium systems is stochastic thermodynamics~\cite{book:stochastic}. By considering Langevin equations or discrete Markov chains, it establishes  trajectorywise energy conservation (leading to the first law of thermodynamics)~\cite{sekimoto98Langevin_thermo,book:sekimoto,PRX2021} and gives a wealth of fluctuation relations (FRs)~\cite{SeifertRPP12}, such as the Jarzynski and Crooks relations~\cite{JarzynskiPRL97,JarzynskiPRE97,CrooksPRE99}. The latter also yields the second law of thermodynamics for the average entropy changes and heat increments. %(with some individual trajectories violating it~\cite{JarzynskiJSP99}). 
%These results are said to be valid arbitrarily far from equilibrium under minimal assumptions. 
Yet, implicitly, all the aforementioned FRs and extended second law of thermodynamics have relied on the existence of a temperature, defined through the ER, to connect microscopic statistics to thermodynamic energy, mechanical and chemical work, and calorimetric heat. 

Several definitions of a nonequilibrium temperature were proposed based on generalizations of the fluctuation-dissipation theorem (FDT) and FRs in active matter~\cite{solon2015pressure,fodor2016EP,NardiniPRX17,MartinPRE21}, systems
with memory~\cite{CuglianoloPRE97,CugliandoloJSTAT05}, and anomalous diffusion~\cite{RainerPRR22,RainerJSTAT12}. Most of these require a thermal bath that obeys the ER.
In this Letter, we adopt an alternative approach, which is not derived from extensions of FRs or FDT, to identify a temperaturelike variable directly, such that the classical-thermodynamic results are recovered.

In some cases, the ER is required to keep the extended FRs connected to physical measurables. For example, the Hatano-Sasa relation~\cite{HatanoPRL01}, an extension of the Jarzynski equality, requires Markovian dynamics only. It therefore relaxes Jarzynski's assumption of a steady-state Boltzmann distribution for a given value of the external protocol~\cite{JarzynskiPRE97}, thus relaxing also the ER requirement. Yet, the effective Hamiltonian and work used by this relation coincide with their actual, measurable  counterparts only under those extra assumptions. Otherwise, they demand knowledge of the complete microstates' steady-state distribution, which is usually inaccessible or requires simplistic models~\cite{RitortPNAS04}.
The temperaturelike variable proposed here, arising directly from thermodynamics, does not rely on those assumptions.

We have two aims: (i) to investigate the fate of existing nonequilibrium FRs and extensions to the second law of thermodynamics once ER is violated;
%the assumption of an ER-based temperature is removed; 
%This also included a new representation of the informatic entropy production [IEP, Eq.~\eqref{eq:rep-EPR}] and a consistent expansion which is independent of the discretization~\cite{SM}. The new expansion clarifies some inconsistencies in previous calculation~\cite{CatesEnt22}.
 and (ii) to formulate a generalized, ER-independent temperature and derive the resulting alternative thermodynamic identities. %which are directly related to thermodynamic measurables. 
%Specifically, without ER: We show that the entropy production~\cite{book:stochastic,SeifertRPP12}, $\Sigma_t$, does not coincide with the thermodynamic heat dissipation, $\cal\diss_t$. (The possibility of such an inconsistency have been proposed in Ref.~\cite{SeifertRPP12}.) By appealing to the Harada-Sasa relation~\cite{HaradaPRL05}, we find that the two quantities are related to the breaking of fluctuation dissipation theorem for two, different perturbations. Additionally, we immediately show that the Crooks FR and Jarzynski equality dictate the statistics of a quantity which may not be identified as dissipated work. 

%%%%%%%%%%%%%%%%%%%%%% To discussion
%The generalized temperature is drive-dependent. It leads to a second-law-like inequality which becomes an equality for flux-free steady states. It is a natural choice in light of the nonnegativity of the mean IEP. As a result, we show that this temperature could be found from calorimetric measurements and path sampling, making it completely model independent. However, its definition implies that the Crooks and Jarzynski relations (requiring beyond-first-moment statistics) \emph{cannot} be established. The connection of the heat to the breaking of FDT via the Harada-Sasa relation is also violated. Potential generalization and extensions of our approach are discussed.
%%%%%%%%%%%%%

\textit{Overdamped stochastic energetics}.---We consider an ensemble of $N$ interacting degrees of freedom (particles), subjected to a deterministic external time-dependent protocol,  $\Lambda_t$.
%which is a prechosen time-varying system parameter. 
The time-dependent stochastic microstate is an $N$-component vector, $\vecX_t$. We couple the particles to an arbitrary medium determining the mobility $\bmu(\vecx,\lambda)$ and diffusivity $\vecD(\vecx,\lambda)$, %which is assumed to provide them with mobility $\bmu(\vecx,\lambda)$ and diffusivity $\vecD(\vecx,\lambda)$ tensors, 
such that the particles follow the overdamped Langevin equation [Eq.~\eqref{eq:langevin} below]; no other property of the surrounding medium is specified. The aforementioned $\vecX_t$ and $\Lambda_t$ denote a particular time-dependent realizations of $\vecx$ and $\lambda$. The particles are subjected to an additional external force $\vecF_\mathrm{ext}(\vecx,\lambda)$, which may be nonconservative and is excluded from the internal energy. We denote the total ($N$-component) force acting on the particles by $\vecF=\vecF_\mathrm{ext}-\boldsymbol{\nabla}H$, where $H(\vecx,\lambda)$ is the system's Hamiltonian. The system is driven out of equilibrium due to the nonconservative force $\vecF_\mathrm{ext}$~\cite{DissipForce} and/or violation of ER [Eq.~\eqref{eq:einstein}]. Thus, the system's steady-state distribution is not Boltzmann, nor can a temperature be conventionally identified.

The system follows the overdamped Langevin equation,
\begin{eqnarray}
\nonumber\dot\vecX_t&=&\bmu(\vecX_t,\Lambda_t)\cdot\vecF(\vecX_t,\Lambda_t) + \grad\cdot\vecD(\vecX_t,\Lambda_t)\\
 && + \vecb(\vecX_t,\Lambda_t)\cdot\noise_t \, ,
 \label{eq:langevin}
\end{eqnarray}
where  $\vecb$ is the noise magnitude satisfying $\vecD=\vecb\cdot\vecb^\T/2$, $\noise_t$ is the normalized thermal noise [$\langle\noise_t\rangle=\veco$, $\langle\noise_t\noise_{t'}\rangle=\vecI\delta(t-t')$, where $\vecI$ is the identity matrix], and the equation is written in the It\^o convention~\cite{book:schuss}. 
The ``spurious drift'' term $\grad\cdot\vecD$ ensures that  Boltzmann distribution is obtained at equilibrium, \ie in systems with a well-defined temperature, Eq.~\eqref{eq:einstein}, under conservative forces~\cite{LubenskyPRE07,CatesEnt22}. The corresponding Fokker-Planck equation~\cite{book:FPE} for the microstates distribution $p(\vecx,t)$ 
is
\begin{eqnarray}
	\label{eq:FPE}
    \frac{\partial p(\vecx,t)}{\partial t} &=& -\grad\cdot\vecJ(\vecx,t)  \, , \\
    \vecJ &=& \bmu\cdot\vecF \, p-\vecD\cdot\grad p \, ,
    \label{eq:flux}
\end{eqnarray}
where $\vecJ$ is the probability flux.

We follow Sekimoto's approach~\cite{sekimoto98Langevin_thermo,book:sekimoto} to identify the internal energy, work, and heat for trajectories in generic Langevin systems. The time derivative of the Hamiltonian is~\cite{book:schuss} 
\begin{equation}
    \dot H_t=\grad H\circ\dot\vecX_t+\frac{\partial H}{\partial\lambda}\dot\Lambda_t \, ,\label{eq:dH}
\end{equation}
where $\circ$ is the Stratonovich product~\cite{STRATO}.
$(\partial H/\partial \lambda)\d\Lambda_t$ is the external work necessary to change the system's internal energy over time $\d t$. The force $\vecF_\mathrm{ext}(\vecX_t,\Lambda_t)$ acting along a displacement $\d\vecX_t$ during $\d t$ performs additional work, $\vecF_\mathrm{ext}\circ\d\vecX_t$.
Thus, the total work rate is
\begin{equation}
    \dot W_t=\frac{\partial H}{\partial\lambda}\dot\Lambda_t+\vecF_\mathrm{ext}\circ\dot\vecX_t \, .
    \label{eq:dW}
\end{equation}
According to the first law of thermodynamics, the heat rate %at time $t$ 
is the remaining energy flow into the system,
\begin{equation}
    \dot Q_t=\dot H_t-\dot W_t=-\vecF\circ\dot\vecX_t \, .\label{eq:dQ}
\end{equation}
Converting the Stratonovich differential into It\^o~\cite{book:schuss} and inserting Eq.~\eqref{eq:langevin}, we find an implicit Langevin equation for the heat in terms of the solutions $\vecX_t$ of Eq.~\eqref{eq:langevin},
\begin{equation}
    \dot Q_t=-\vecF\cdot\bmu\cdot\vecF-\grad\cdot(\vecD\cdot\vecF) -\vecF\cdot\vecb\cdot\noise_t \, .\label{eq:dQlangevin}
\end{equation}

The above is a natural adaptation of the internal energy, work, and heat to the microscale~\cite{sekimoto98Langevin_thermo,book:sekimoto}. A similar extension of the system's entropy is less trivial. %It is common to 
Consider the instantaneous nonequilibrium stochastic entropy~\cite{book:stochastic,SeifertRPP12},
$S(\vecx,t)=-\ln p(\vecx,t)$,
whose ensemble average is the  Shannon entropy~\cite{NONEQENT,Shannon1948,BOOK:infotheor}, $\langle S\rangle=-\int \d\vecx\, p(\vecx,t)\ln p(\vecx,t)$.
The relation between classical-thermodynamic entropy and stochastic entropy is established for equilibrium states by finding the distribution $p$ that maximizes %the Shannon entropy 
$\langle S\rangle$~\cite{book:kardar1} (for a  fixed protocol value $\lambda$). Combining the stochastic chain rule~\cite{book:schuss} and Eqs.~\eqref{eq:langevin}--\eqref{eq:flux} yields the time derivative of $S_t=S(\vecX_t,t)$,
\begin{equation}
    \dot S_t=\grad\cdot\left(\frac{\vecJ}{p}\right)-\frac{1}{p}\grad\cdot(\vecD\cdot\grad p) - [(\grad\ln p)\cdot\vecb]\cdot\noise_t \, , \label{eq:dSlangevin}
\end{equation}
where the last two terms are often omitted~\cite{SeifertPRL05} as they vanish on average.

If a well-defined temperature $T$ exists, it is instructive to consider the thermodynamic heat dissipation, $\kb T R_t$. It quantifies how much extra calorimetric heat $\dot Q_t$ is wasted in an  irreversible process. 
According to the second law of thermodynamics, the average thermodynamic dissipation $\langle \dot\diss_t\rangle$ is non-negative~\cite{JarzynskiJSP99},
\begin{equation}
    \dot \diss_t\equiv\dot S_t-\frac{\dot Q_t}{\kb T}\,,\qquad\langle\dot \diss_t\rangle\geq0\,.\label{eq:2nd_law}
\end{equation}
Recall that $\kb T\dot S_t$ is the reversible contribution to the heat.
Our goal is to arrive at a similar relation in the absence of the conventional temperature $T$.

\textit{Informatic entropy production (IEP)}.---The IEP, $\Sigma_t$, quantifies the breaking of time-reversal symmetry in the microstates' dynamics~\cite{book:stochastic}. 
Consider a discretized trajectory drawn from Eq.~\eqref{eq:langevin}, $(\vecX_0,\dots,\vecX_{K \Delta t})=(\vecx_0, \dots, \vecx_K)$, with arbitrarily small time  step $\Delta t$. The probability distribution of obtaining this trajectory under the  protocol sequence $(\lambda_0,\lambda_1,\ldots,\lambda_K)$, $\rights\calP\equiv\Pr(\vecx_0,\vecx_1,\ldots,\vecx_K)$, 
is compared with %its reversed counterpart\,---\,
the probability distribution of obtaining the reverted trajectory, $\lefts\calP\equiv\Pr(\vecx_K,\vecx_{K-1},\ldots,\vecx_{0})$, under the reverted protocol $(\lambda_{K},\lambda_{K-1},\ldots,\lambda_{0})$. Explicitly~\cite{SeifertRPP12,book:stochastic},
\begin{equation}
    \Sigma_t=\ln\frac{\rights\calP}{\lefts\calP} \, .\label{eq:def-EP}
\end{equation}

For Markovian systems [such as Eq.~\eqref{eq:langevin}], it is helpful to decompose the trajectory probabilities using the infinitesimal-time propagator, $\Pr(\vecx_{k+1},\lambda_{k+1}|\vecx_k,\lambda_k)$, as $\rights\calP=p(\vecx_0,0)\prod_{k=1}^K\Pr(\vecx_{k},\lambda_{k}|\vecx_{k-1},\lambda_{k-1})$. This decomposes the IEP rate at time $t=k\Delta t$ into
\begin{eqnarray}
    \nonumber\dot\Sigma_t&=&\dot S_t+\dot\Omega_t \, ,\\
    \dot\Omega_t&=&\lim_{\Delta t\to0}\frac{1}{\Delta t}\ln\frac{\Pr(\vecx_{k+1}\, , \lambda_{k+1}|\vecx_k,\lambda_k)}{\Pr(\vecx_k,\lambda_k|\vecx_{k+1},\lambda_{k+1})} \, .\label{eq:def-infoQ}
\end{eqnarray}
In Eq.~\eqref{eq:def-infoQ} we  identify the time derivative of the entropy, $\dot S_t=\ln[p(\vecx_{k+1},(k+1)\Delta t)/p(\vecx_{k},k\Delta t)]/\Delta t$, given by Eq.~\eqref{eq:dSlangevin}. 
%which is related to the Langevin system through Eq.~\eqref{eq:dSlangevin}. 
We  refer to the remainder, $\Omega_t$, as the ``informatic heat" (IH). The propagator in the denominator, $\Pr(\vecx_k,\lambda_k|\vecx_{k+1},\lambda_{k+1})$, is the exact same propagator of Eq.~\eqref{eq:langevin}, where the jump process is reverted [\ie the process starts at $\vecx_{k+1}$ at time $t=k\Delta t$ and ends up at $\vecx_k$ at time $(k+1)\Delta t$ under the  protocol value $\lambda_{k+1}$]. 

Focusing on the IEP has two advantages. First, its ensemble average $\langle\Sigma_t\rangle$ is a Kullback-Leibler divergence;
%(from the reversed-trajectory distribution to the forward-trajectory one), 
 thus, it is non-negative~\cite{BOOK:infotheor}. 
%and thus guaranteed to be nonnegative~\cite{BOOK:infotheor}. 
Second, the dissipation coincides with the IEP, $\dot\diss_t=\dot\Sigma_t$, in a wide range of scenarios satisfying the Einstein relation~\cite{book:stochastic}.  In particular, one finds $\dot Q_t=-\kb T\dot\Omega_t$~\cite{CatesEnt22} (hence the term informatic heat for $\Omega_t$). For these cases, therefore, one does not have to resort to simplistic models for the dynamics and then employ Eq.~\eqref{eq:dQlangevin} to obtain $\dot Q_t$. %the thermodynamic heat rate. 
Instead, one can access $\dot Q_t$ %the thermodynamic heat rate 
from the statistics of trajectories [Eqs.~\eqref{eq:def-EP} and~\eqref{eq:def-infoQ}] without any prior knowledge of the underlying dynamics~\cite{BiskerNATCOM19,dunkelPRL21}.  %equations of motion~\cite{BiskerNATCOM19,dunkelPRL21}. 
Clearly, the correspondences $\dot\Omega_t$--$\dot Q_t$ and $\dot\Sigma_t$--$\diss_t$ assume the ER with its well-defined temperature.

More specifically, in Ref.~\cite{JSTAT24} the IH rate is found to satisfy~\cite{RefCatesFoot}
\begin{equation}
  \dot\Omega_t = [(\bmu\cdot\vecF)\cdot\vecD^{-1}]\circ\dot\vecX_t.\label{eq:IHrate}   
\end{equation}
Therefore, if $\vecD$ and $\bmu$ are not related by a scalar, the informatic $\dot\Sigma_t$ is no longer related to the measurable $\dot \diss_t$, even on average. Consequently, the non-negativity of $\langle\dot \diss_t\rangle$ (\ie the second law) can no longer be established based on trajectory statistics. 
Below we propose a definition of a nonequilibrium temperature that mends this discrepancy.

\textit{Nonequilibrium temperature}.---We return to the second law, Eq.~\eqref{eq:2nd_law}. Noting 
that it concerns only the means of Eqs.~\eqref{eq:dQlangevin} and~\eqref{eq:dSlangevin}, we average over the noise realizations (while using $\langle\noise_t\rangle=\veco$), and over the instantaneous position, where $\langle[\cdot](\vecX_t,\Lambda_t)\rangle=\int \d\vecx\, p(\vecx,t)[\cdot](\vecx,\Lambda_t)$.
Integration by parts gives $\langle \dot Q_t\rangle=-\int \d\vecx\, \vecJ\cdot\vecF$ and $\langle \dot S_t\rangle=-\int \d\vecx\,\vecJ\cdot\grad \ln p$. Let us posit that the second law of thermodynamics, Eq.~\eqref{eq:2nd_law}, still holds, but with a nonequilibrium temperature $\teff$. Using the expressions for $\langle \dot Q_t\rangle$ and $\langle \dot S_t\rangle$, we express the average dissipation rate of Eq.~\eqref{eq:2nd_law} as
\begin{equation}
    \langle \dot \diss_t\rangle=\int \d\vecx\, \frac{\vecJ}{p}\cdot\vecD^{-1}\cdot\left(\frac{\vecD}{\kb \teff}\cdot\vecF p-\vecD\cdot\grad p\right).\label{eq:diss_noneq}
\end{equation}
The term in the parentheses, which depends on our choice of $\teff$, is reminiscent of $\vecJ$ [Eq.~\eqref{eq:flux}]. 
%This separates out in parentheses a term reminiscent of the probability flux $\vecJ$, which depends on our choice of $\teff$. 

We now impose the following physical conditions on  Eq.~\eqref{eq:diss_noneq}: (i) To satisfy the second law, $\langle \dot \diss_t\rangle$ should be a non-negative functional of $\vecJ$. (ii) To agree with Onsager's theory~\cite{onsagerPR31b,onsagerPR31a,BOOK:Ottinger}, the right-hand side should reduce to a quadratic form of $\vecJ$ near equilibrium and turn  the second-law inequality into an equality for reversible processes ($\vecJ\to \veco$). (iii) The variable $\teff$ should be a configuration-independent scalar. The simplest choice for the dissipation functional that satisfies these three conditions is 
    $\langle\dot \diss_t\rangle=\int \d\vecx (\vecJ\cdot\vecD^{-1}\cdot\vecJ)/p$~\cite{SeifertFOOT}.
To complete our construction, we solve the equation $\langle\dot \diss_t\rangle=\langle\dot S_t\rangle-\langle\dot Q_t\rangle/(\kb \teff)$ for $\kb\teff$, and find the following drive-dependent generalized temperature:
\begin{equation}
    \kb \teff=\frac{\int \d\vecx\,\vecJ\cdot\vecF}{\int \d\vecx\,\vecJ\cdot\vecD^{-1}\cdot(\bmu\cdot\vecF)} \, .\label{eq:noneqtemp}
\end{equation}
We have thus recovered the second law of thermodynamics  for general Markovian overdamped Langevin dynamics (Eq.~\eqref{eq:langevin}).

Equation~\eqref{eq:noneqtemp} is our central result. If ER holds, such that $\vecD^{-1}\cdot\bmu=(\kb T)^{-1}$, we readily get $\teff=T$. Otherwise, the generalized temperature is an intricate quantity. Computing it directly from Eq.~\eqref{eq:noneqtemp} is a challenge. Below we propose how to estimate it from experiments (see, e.g., Eq.~\eqref{eq:noneqtemp_ratio}). This is expected for nonequilibrium systems with arbitrarily complex $\vecJ$, $\vecD$, $\bmu$, and $\vecF$. 
%Being drive-dependent, it is evidently less universal than the conventional temperature. 
%This is to be expected for nonequilibrium systems with arbitrarily complex $\vecJ$, $\vecD$, $\bmu$, and $\vecF$. 
Furthermore, $\teff$  depends on time via the protocol $\Lambda_t$ affecting all these quantities. Even for a constant protocol $\Lambda_t=\mathrm{const}$, $\teff$ will change in time due to the instantaneous distribution $p(\vecx,t)$ entering $\vecJ$.  Indeed, nonequilibrium steady states generally have $\vecJ\ne\veco$, as nonconservative forces %activity 
and drift create persistent fluxes across scales. Despite these caveats, Eq.~\eqref{eq:noneqtemp} should play an experimentally meaningful role as a generalized temperature in nonequilibrium thermodynamics~\cite{GranickSOFT20,CapriniCP24,DimaYael24,GalorYael24}, as discussed below.
%(through $\vecD(\vecx,\lambda(t))$, $\bmu(\vecx,\lambda(t))$, and $\vecF(\vecx,\lambda(t))$, which also appear in $\vecJ(\vecx,t)$) and also explicitly (through the instantaneous distribution $p(\vecx,t)$ appearing in $\vecJ(\vecx,t)$). Thus, even with a constant protocol $\lambda$, the temperature may change along a trajectory that approached the steady state at that $\lambda$. While having a forcing-dependent temperature may not seem universal (as one expects from equilibrium, classical thermodynamics), the nonequilibrium limit may take this intuition away, and a well-constructed temperature may just as well depend on the arbitrarily-complex $\vecJ$, $\vecD$, $\bmu$, and $\vecF$. Indeed, more often than not, nonequilibrium steady states may have $\vecJ\ne\veco$, as activity or drift lead to eternal fluxes across scales. Equation~\eqref{eq:noneqtemp} may, hopefully, be useful in this limit (as discussed later in the letter). 

\textit{Consequences}.---First, we reestablish the connection between the IH and heat and between the IEP and dissipation, \ie between information and thermodynamics. Substituting Eq.~\eqref{eq:langevin} in Eq.~\eqref{eq:IHrate}, we find
\begin{eqnarray}
    \nonumber\dot\Omega_t&=&(\bmu\cdot\vecF)\cdot\vecD^{-1}\cdot(\bmu\cdot\vecF)+\grad\cdot(\bmu\cdot\vecF)\\
    &&+[(\bmu\cdot\vecF)\cdot\vecD^{-1}\cdot\vecb]\cdot\noise_t \,,\label{eq:dOMEGAlangevin}
\end{eqnarray}
which, combined with Eq.~\eqref{eq:dSlangevin} and the relation $\dot\Sigma_t=\dot S_t+\dot\Omega_t$, gives the IEP rate as (cf. Ref.~\cite{PigolottiPRL17})
\begin{equation}
    \dot\Sigma_t=\frac{\vecJ}{p}\cdot\vecD^{-1}\cdot\frac{\vecJ}{p}+\frac{2}{p}\grad\cdot\vecJ +\left(\frac{\vecJ}{p}\cdot\vecD^{-1}\cdot\vecb\right)\cdot\noise_t \, .\label{eq:dSIGMAlangevin}
\end{equation}
Upon taking the averages of Eqs.~\eqref{eq:dOMEGAlangevin} and~\eqref{eq:dSIGMAlangevin}, we obtain
\begin{equation}
    \kb \teff= -\langle \dot Q_t\rangle / \langle \dot \Omega_t\rangle \, ,\label{eq:noneqtemp_ratio}
\end{equation}
and 
\begin{equation}
    \langle\dot \diss_t\rangle=\langle\dot \Sigma_t\rangle=\int \d\vecx\left(\vecJ\cdot\vecD^{-1}\cdot\vecJ\right) / p \,\geq\,0 \, .\label{eq:quad}
\end{equation}

Note that the correspondence holds for the means only and not for the full statistics, $\dot Q_t\ne-\kb\teff\dot \Omega_t$ and $\dot \diss_t\ne\dot \Sigma_t$. For example, the variances of the former are different,  $\langle(\d\Omega_t)^2\rangle_\ccc=2\d t\int (\bmu\cdot\vecF)\cdot\vecD\cdot(\bmu\cdot\vecF)p\d \vecx$, whereas $\langle (\d Q_t)^2\rangle_\ccc=2\d t\int\vecF\cdot\vecD\cdot\vecF p\d \vecx\neq(\kb\teff)^2\langle (\d\Omega_t)^2\rangle_\ccc$. (The subscript $\ccc$ denotes a cumulant.) Since the Jarzynski equality~\cite{JarzynskiPRL97,JarzynskiPRE97} and Crooks fluctuation theorem~\cite{CrooksPRE99} require statistics beyond the mean, they are not restored by the generalized $\teff$. Thus, without the ER, the various FRs~\cite{SeifertRPP12} remain related only to informatic quantities (the IEP and IH) \cite{HatanoPRL01} and not to thermodynamics (dissipation and heat). 

Another relation to be examined is the one between the heat of stochastic systems and the departure from FDT. Such a connection was established by Harada and Sasa~\cite{HaradaPRL05}, implicitly assuming the ER and a configuration-independent scalar mobility. (We recall that the ER and FDT are equivalent only in equilibrium~\cite{BaiesiJSP2009}.) Connections found recently for various active systems between FDT violation and the IH (\eg Refs.~\cite{NardiniPRX17,MartinPRE21}) have similarly made these two strong assumptions.
As discussed above, without these assumptions the IH and the thermodynamic heat are separate quantities. In addition, the assumptions underlying FDT do not apply~\cite{SM}. 
It is then unclear which heat is connected to FDT violation---the IH or the thermodynamic heat? 
We find that neither connection can be established in the general case~\cite{SM}, implying that the Harada-Sasa relation breaks down as well.

Nevertheless, generalized versions of the FDT and the Harada-Sasa relation can be derived in the absence of the ER. Here we outline these results; see the Supplemental Material for details~\cite{SM}. Instead of considering the response to an impulsive linear potential gradient as was done in Ref.~\cite{HaradaPRL05}, we consider a perturbation of the force of the form $\delta\vecF(\vecx,s)=-\delta h_\mu(s)\bmu^{-1}(\vecx,s)\cdot\vecD(\vecx,s)\cdot\grad x_\mu$ at time $s<t$, in response to the external perturbation $\delta\vech(s)$.
The difference between the response of the average force to such perturbation and its correlation with the position are found to be related to the thermodynamic heat~\cite{SM},
\begin{eqnarray}
    \nonumber\langle\dot Q_s\rangle\!=\!-\!\lim_{t\to s^+} \! \left(\frac{\d}{\d s}\langle\vecF(\vecX_t,t)\cdot\vecX_s\rangle +\frac{\delta}{\delta\vech(s)}\cdot\langle \vecF(\vecX_t,t)\rangle\right) \,.\\  \label{eq:HSR_heat}
\end{eqnarray}
If, instead of $\vecF$, we consider the observable  $\vecD^{-1}\cdot\bmu\cdot\vecF$, we obtain a similar relation for the IH,
\begin{multline}
  \langle\dot\Omega_s\rangle = \lim_{t\to s^+}\left(\frac{\d}{\d s}\langle[(\bmu\cdot\vecF)\cdot\vecD^{-1}](\vecX_t,t)\cdot\vecX_s\rangle\right.\\\left.+\frac{\delta}{\delta\vech(s)}\cdot\langle [(\bmu\cdot\vecF)\cdot\vecD^{-1}](\vecX_t,t)\rangle\right)\,.\label{eq:HSR_IH}
\end{multline}
Thus the IH and thermodynamic heat are related to violations of a generalized FDT for different observables. 
%It is important to note that the perturbation $\delta\vecF$ chosen above may be very hard to realize in experiment, as it requires knowing $\vecD$ and $\bmu$ in full detail. 
If the ER holds, the two observables coincide and the classical FDT is recovered. With the additional assumption of a configuration-independent scalar mobility the Harada-Sasa relation is recovered as well~\cite{SM}.

%A further note on the relation between information and thermodynamics concerns the effect of coarse-graining on the second law [Eq.~\eqref{eq:quad}]. 
%The second law $\langle\dot \diss_t\rangle\geq0$ immediately follows from it, with equality at flux-free ($\vecJ=\veco$) steady states (that is, equilibria, up to the abovementioned unusual circumstances). On the other hand, nonequilibrium systems, typically having nonzero flux ($\vecJ\neq\veco$), follow the inequality even at steady state. 
%Complete knowledge of the IEP would make Eq.~\eqref{eq:quad} an equality. 
%For any nonzero flux the relation is an inequality. 
%Reliably inferring the IEP, however, is an ongoing challenge. Since the average IEP is a Kullback-Liebler divergence, according to the data processing inequality~\cite{BOOK:infotheor} any coarse-graining of the coordinates would push its lower bound upward. Equivalently, computing the flux $\tilde\vecJ$ of a smaller set of variables $\tilde\vecx$, and estimating their covariance rate (diffusion coefficient) $\tilde\vecD$, would yield a lower bound for the dissipation, $\langle \dot \diss_t\rangle\geq\int \d\tilde\vecx \tilde\vecJ\cdot\tilde\vecD^{-1}\cdot\tilde\vecJ/\tilde p\geq0$. This is reminiscent of the thermodynamic uncertainty relations~\cite{GingrichPRL16,SeifertPRL15}.
%{\color{olive}[There is a confusion here between the IEP and its rate. I remember discussing it but now I am confused again.]}

We now revisit additional classical-thermodynamic results demonstrating the relevance of the generalized temperature $\teff$.  More details regarding these consequences are found in the Supplemental Material~\cite{SM}. (i) Consider a process that starts and ends at the same protocol value, $\Lambda_{t_\mathrm{f}}=\Lambda_{0}$, and then allowed to relax to steady state. Since $\oint_0^\infty \d t \langle\dot{S}_t\rangle=0$, we obtain the Clausius inequality, $\oint_0^\infty \d t\langle\dot Q_t\rangle/(\kb\teff)=-\oint_0^\infty \d t\langle\dot \Sigma_t\rangle\leq0$~\cite{SM}. (ii) Construct a Carnot-like cyclic engine consisting of an ``isothermal'' process (with constant $\teff_\mathrm{H}$ and $\langle Q_\mathrm{in}\rangle>0$), an ``adiabatic'' process (in which the system ``cools'' down to $\teff_\mathrm{C}$), followed by another isothermal process (with constant $\teff_\mathrm{C}$, and $\langle Q_\mathrm{out}\rangle<0$), and an adiabatic process (where the system reaches $\teff_\mathrm{H}$). The efficiency of such an engine is found as $\langle-W_\mathrm{cycle}\rangle/\langle Q_\mathrm{in}\rangle = 1-\teff_\mathrm{C}/\teff_\mathrm{H}+\kb\teff_\mathrm{C}\langle\Sigma_\mathrm{cycle}\rangle/\langle Q_\mathrm{in}\rangle$, where $\langle\Sigma_\mathrm{cycle}\rangle$ is the mean IEP during a cycle~\cite{SM}. Thus, the mean IEP quantifies the cycle's irreversibility compared to a generalized Carnot efficiency, $1-\teff_\mathrm{C}/\teff_\mathrm{H}$. (See also Ref.~\cite{DattaPRX22} in the context of nonequilibrium engines.) (iii) Define the nonequilibrium free energy, $A_t=H_t-\kb\teff S_t$.  It is easy to show that it bounds the extractable work in isothermal (constant $\teff>0$) transformations, $\langle-\dot W_t\rangle=-(\langle \dot A_t\rangle+\kb\teff\langle\dot\Sigma_t\rangle)\leq-\langle \dot A_t\rangle$~\cite{SM}. 

An interesting consequence of Eq.~\eqref{eq:noneqtemp} relates to a particular class of nonequilibrium systems having multiple ``temperatures'', \eg a mixture of different types of particles, each of which is  subjected to a different ``thermostat''~\cite{JoannyPRE15,FreyPRL16}. These models may be relevant, for instance, for mixtures of passive and active particles~\cite{CatesPRL15,ArielPRL15,MarkovichBIO23,ArielJSP15,ArielPHYSA19}. They yield a temperaturelike scalar that is a weighted average of the thermostats' temperatures. Considering a diagonal but nonscalar $\vecD^{-1}\cdot\bmu$, with different inverse temperatures as its eigenvalues, we may view Eq.~\eqref{eq:noneqtemp} as a generalization of such a weighted average. These systems may attain $\vecJ=\veco$, which results in a ``$0/0$'' ambiguity for $\teff$. However, perturbing different global parameters amounts to taking a perturbation $\vecJ$ along different axes, thus picking up different eigenvalues (inverse temperatures) of $\vecD^{-1}\cdot\bmu$. We show a simplistic example in the Supplemental Material~\cite{SM}.

\textit{Conclusion}.---We have analyzed a generic nonequilibrium Langevin system [Eq.~\eqref{eq:langevin}] that violates the ER [Eq.~\eqref{eq:einstein}]. This describes ubiquitous scenarios in driven and active materials since the relation between dissipation (mobility) and fluctuations (diffusivity) is a particular consequence of the equilibrium Boltzmann distribution~\cite{KuboRPP66}. We have demonstrated the far-reaching consequences of ER violation, which may be summarized as the detachment of informatic quantities (IH, IEP) from their physical thermodynamic counterparts (heat, dissipation) and the resulting invalidation of fluctuation theorems which rely on these correspondences.

To restore the information-thermodynamics correspondence we have imposed the second law of thermodynamics directly and identified a generalized temperature $\teff$ [Eq.~\eqref{eq:noneqtemp}], which gives all the thermodynamically required properties: non-negative dissipation, zero dissipation at equilibrium, linear thermodynamic stability, and the temperature being a macroscopic (integrated) property. 
The generalized temperature is model-independent---with a calorimetric measurement of the heat and path sampling for the IH, one may find the system's $\teff$ with no required knowledge of the underlying dynamics.
In particular, the variety of tools available for the estimation of IEP from trajectories (e.g., sampling marginal distributions~\cite{BiskerJSM2017,Parrondo07EPR,dunkelPRL21} or the thermodynamic uncertainty relations~\cite{GingrichPRL16,SeifertPRL15}) will be useful, since in our construction the average IEP remains connected to the average thermodynamic dissipation.
The generalized temperature does not restore the various FRs~\cite{JarzynskiPRL97,CrooksPRE99,SeifertRPP12,HaradaPRL05}, which in the absence of the ER no longer dictate the statistics of thermodynamic properties~\cite{HatanoPRL01}. 

The three conditions that have been used to find $\teff$, Eq.~\eqref{eq:noneqtemp}, do not uniquely determine it. Within linear irreversible thermodynamics~\cite{onsagerPR31a,onsagerPR31b,BOOK:Ottinger}, they do guarantee a single generalized temperature, as they impose the exact leading $\mathcal{O}(J^2)$ term, $\langle\dot \diss_t\rangle=\int \d\vecx (\vecJ\cdot\vecD^{-1}\cdot\vecJ)/p$.
This leading term is the complete dissipation functional when the Einstein relation holds.
If one additionally requires this dissipation functional beyond linear theory, $\teff$ is then uniquely determined by Eq.~\eqref{eq:noneqtemp}. The fact that $\teff$ is the ratio between $\langle \dot Q_t\rangle$ (thermodynamics) and $\langle\dot\Omega_t\rangle$ (information) is then a useful consequence of this choice of dissipation functional.

The temperature $\teff$ obtained here is a property of a single closed system exchanging heat with an arbitrary bath. 
It dictates the direction of heat flow into or out of the system based on the change in the system's entropy.
The fact that the entropy production is non-negative, $\langle\dot \diss_t\rangle \geq0$, implies a positive $\teff$ if heat is leaving the system and a negative one if heat is entering the system at the nonequilibrium steady state. At equilibrium, there is no heat flow.
It remains to be explored whether $\teff$ also dictates the direction of heat flow between two systems in contact.

The applicability of $\teff$ as a nonequilibrium temperature can be put to test in simple experimental scenarios. For example, one can make the mobility and diffusivity separately isotropic or anisotropic by controlling, respectively, the particle shape and athermal randomized forces. This will break the ER and allow the tuning of $\teff$ according to Eq.~\eqref{eq:noneqtemp}. One can then study various manifestations of the second law based on $\teff$ and its spatial variation~\cite{GalorYael24}.

\begin{acknowledgments}
We greatly benefited from discussions with David Dean and Yael Roichman. T.M. acknowledges funding from the Israel Science Foundation (Grant No. 1356/22). G.A. and H.D. acknowledge funding from the Israel Science Foundation (Grant No. 1611/24).
\end{acknowledgments}

%\bibliography{PR23BIB}

%%%%%%%%%%%%%%%%%%%%%%%%%%%%%%%%%%%%%%%%%%%%

%%%%%%%%%%%%%%%%%%%%%%%%%%%%%%%%%%%%%%%%%%%%
\renewcommand{\theequation}{S\arabic{equation}}
\setcounter{equation}{0}
\renewcommand{\thefigure}{S\arabic{figure}}
\setcounter{figure}{0}
\renewcommand{\thetable}{S\arabic{table}}
\setcounter{table}{0}
\renewcommand{\thesection}{S\Roman{section}}
\setcounter{section}{0}

\begin{widetext}
\newpage

\section*{Supplementary Material: Second Law of Thermodynamics without Einstein Relation}

\section{Review: Response and correlation functions}\label{sec:old}

In this section, we recall the classical fluctuation-dissipation theorem~\cite{KuboRPP66} and the Harada-Sasa relation~\cite{HaradaPRL05}. We do so in order to  highlight the assumptions that underlie both equations and demonstrate the importance of the Einstein relation. All results presented in this section are known; we briefly repeat them for the sake of the sections that follow.

\subsection{The Fokker-Planck equation}\label{sec:FPE}

In Eq.~(3) of the main text we consider the Fokker-Planck equation,
\begin{equation}
    \frac{\partial p(\vecx,t)}{\partial t}=-\grad\cdot[\bmu(\vecx,t)\cdot\vecF(\vecx,t)p(\vecx,t)-\vecD(\vecx,t)\cdot\grad p(\vecx,t)].\label{eq:FPE}
\end{equation}
Numerous authors have discussed why $\grad\cdot[\vecD(x,t)\cdot\grad]$ is the physically-relevant order of derivatives; see, \eg Refs.~\cite{book:schuss,LubenskyPRE07,CatesEnt22,JSTAT24}. In short, assuming the Einstein relation is satisfied, $\vecD(\vecx)=\kt\bmu(\vecx)$, and the forces are conservative, $\vecF(\vecx)=-\grad H(\vecx)$, this order of derivatives is the one that yields the Boltzmann factor, $p(\vecx,t\to\infty)\sim e^{-H(\vecx)/(k_\mathrm{B}T)}$, as the steady-state (equilibrium) distribution of the Fokker-Planck equation. The spurious drift term that appears in Eq. (2) of the main text was included in order to obtain the corresponding Langevin equation in the It\^o representation~\cite{LubenskyPRE07}.

For later purposes, it will be convenient to define the Fokker-Planck operator $\lag$,
\begin{equation}
    \frac{\partial p(\vecx,t)}{\partial t}=\lag(\vecx,t) p(\vecx,t),\qquad\lag(\vecx,t)\equiv-\grad\cdot[\bmu(\vecx,t)\cdot\vecF(\vecx,t)-\vecD(\vecx,t)\cdot\grad].\label{eq:FPoperator}
\end{equation}
Its Hermitian conjugate is
\begin{equation}
    \lag^\dagger(\vecx,t)\equiv[\bmu(\vecx,t)\cdot\vecF(\vecx,t)+\grad\cdot\vecD(\vecx,t)]\cdot\grad.\label{eq:FPadjoint}
\end{equation}
In Eq.~(4) of the main text, we identified the probability flux $\vecJ$,
\begin{equation}
    \frac{\partial p(\vecx,t)}{\partial t}=-\grad\cdot\vecJ(\vecx,t),\qquad\vecJ(\vecx,t)\equiv\bmu(\vecx,t)\cdot\vecF(\vecx,t)p(\vecx,t)-\vecD(\vecx,t)\cdot\grad p(\vecx,t).\label{eq:FPflux}
\end{equation}

\subsection{The fluctuation-dissipation theorem} \label{sec:FDTwE}

All derivations in this Section are based on Ref.~\cite{book:zwanzig}. (See also Refs.~\cite{FDTderiv,book:FPE}.) We may formally write the Fokker-Planck propagator as
\begin{equation}
    \Pr(\vecx,t|\vecy,s)=\exp\left[\int_{s}^{t}\d t'\lag(\vecx,t')\right]\delta(\vecx-\vecy).
\end{equation}
With this, we can compute, e.g., the autocorrelation between two configuration-dependent variables, $A(\vecX_t,t)$ and $B(\vecX_s,s)$,
\begin{equation}
    \langle A(\vecX_t,t)B(\vecX_s,s)\rangle=\int \d\vecy p(\vecy,s)B(\vecy,s)\exp\left[\int_{s}^{t}\d t'\lag^\dagger(\vecy,t')\right]A(\vecy,t).\label{eq:ABcorr}
\end{equation}
% We can also compute the mean of $A(\vecX_t,t)$ given the distribution at an earlier time, $p(\vecy,s)$,
% \begin{equation}
%     \langle A(\vecX_t,t)\rangle=\int \d\vecy p(\vecy,s)\exp\left[\int_{s}^{t}\d t'\lag^\dagger(\vecy,t')\right]A(\vecy,t).\label{eq:Amean}
% \end{equation}

Recall the following identity under the variation of the operator $\lag\to\lag+\delta\lag$:
\begin{equation}
\delta\exp\left[\int_s^t\d t'\lag(\vecx,t')\right]=\int_s^t\d s'\exp\left[\int_{s'}^t\d t''\lag(\vecx,t'')\right]\delta\lag(\vecx,s')\exp\left[\int_s^{s'}\d t'\lag(\vecx,t')\right].\label{eq:FPOvar}
\end{equation}
Its Hermitian conjugate reads
\begin{equation}
    \delta\exp\left[\int_s^t\d t'\lag^\dagger(\vecx,t')\right]=\int_s^t\d s'\exp\left[\int^{s'}_{s}\d t'\lag^\dagger(\vecx,t')\right]\delta\lag^\dagger(\vecx,s')\exp\left[\int_{s'}^{t}\d t''\lag^\dagger(\vecx,t'')\right].\label{eq:FPADJvar}
\end{equation}
From Eq.~\eqref{eq:FPADJvar}, one finds for a perturbation $\delta L(x,s') = -\delta s \delta(s'-s) L(x,s')$
\begin{equation}
    \frac{\d}{\d s}\exp\left[\int_s^t\d t'\lag^\dagger(\vecx,t')\right]=-\lag^\dagger(\vecx,s)\exp\left[\int_s^t\d t'\lag^\dagger(\vecx,t')\right].\label{eq:FPAsdev}
\end{equation}

With this, we proceed to take the $s$-derivative of Eq.~\eqref{eq:ABcorr}. Using Eq.~\eqref{eq:FPAsdev}, integrating by parts, and utilizing Eqs.~\eqref{eq:FPoperator}\,--\,\eqref{eq:FPflux}, we find that
\begin{eqnarray}
    \frac{\d}{\d s}\langle A(\vecX_t,t)B(\vecX_s,s)\rangle&=&\langle A(\vecX_t,t)\left.\frac{\partial B}{\partial t}\right|_{\vecX_s,s}\rangle+\int\d\vecx \vecJ(\vecx,s)\cdot[\grad B(\vecx,s)]\exp\left[\int_s^t\d t'\lag^\dagger(\vecx,t')\right]A(\vecx,t)\nonumber\\&&+\int \d\vecx p(\vecx,s)[\vecD(\vecx,s)\cdot\grad B(\vecx,s)]\cdot\grad\left\{\exp\left[\int_s^t\d t'\lag^\dagger(\vecx,t')\right]A(\vecx,t)\right\}.\label{eq:ABdev}
\end{eqnarray}
Now, introduce an instantaneous perturbation to the system's Hamiltonian, $H(\vecx,s)\to H(\vecx,s)+\delta h_H(s)\pertH(\vecx)$, where $\delta h_H(s)$ controls the timing of this impulsive change and $\pertH(\vecx)$ is the actual change to the potential. As a result, the adjoint Fokker-Planck operator [Eq.~\eqref{eq:FPadjoint}] is modified as $\delta\lag^\dagger(\vecx,s)=\delta h_H(s)[-\bmu(\vecx,s)\cdot\grad\pertH(\vecx)]\cdot\grad$. Thus, using Eq.~\eqref{eq:FPADJvar}, the response of the mean value of the observable $A(\vecX_t,t)$ to the perturbation is
\begin{equation}
    \frac{\delta\langle A(\vecX_t,t)\rangle}{\delta h_H(s)}=-\int\d\vecx p(\vecx,s)[\bmu(\vecx,s)\cdot\grad\pertH(\vecx)]\cdot\grad\left\{\exp\left[\int_{s}^{t}\d t'\lag^\dagger(\vecx,t')\right]A(\vecx,t)\right\}.\label{eq:resPOTENT}
\end{equation}

In order to obtain the classical fluctuation-dissipation theorem, we now assume that the system was in equilibrium at the time of the perturbation, $s$. ``Equilibrium'', in fact, implies two distinct assumptions: (1) that Einstein relation holds, $\vecD(\vecx,s)=\kt(s)\bmu(\vecx,s)$, where $\kt(s)$ is a uniform scalar temperature, and (2) that the system was at a flux-free steady-state, $\vecJ(\vecx,s)=\veco$. Upon inserting these assumptions in Eq.~\eqref{eq:ABdev} (with $B=\tilde H$) and comparing with Eq.~\eqref{eq:resPOTENT}, we obtain the equilibrium fluctuation-dissipation theorem~\cite{KuboRPP66,book:zwanzig}:
\begin{equation}
    \frac{\d}{\d s}\langle A(\vecX_t,t)\pertH(\vecX_s)\rangle=-\kt\frac{\delta\langle A(\vecX_t,t)\rangle}{\delta h_H(s)}.\label{eq:FDT}
\end{equation}

Important to this work is the case in which Einstein relation is invalid at $s$ 
(due to, \eg activity), so there is no uniform scalar $\kt(s)$ relating $\vecD(\vecx,s)$ to $\bmu(\vecx,s)$. This means that the response to a perturbed potential [Eq.~\eqref{eq:resPOTENT}] may not be related to the correlation with this potential [Eq.~\eqref{eq:ABdev}]. Therefore, the fluctuation-dissipation theorem [Eq.~\eqref{eq:FDT}] is not satisfied. This is hinted by the fact that the correlation and response functions differ in units by a factor of energy units.

%We finally note that typically, one may find Eq.~\eqref{eq:resPOTENT} in the literature with the time-derivative $\d/\d t$ rather than $\d/\d s$~\cite{KuboRPP66,FDTderiv}. This is because time-independent systems that are at steady state at $t>s$ as well are considered, and of course in this limit the two derivative are equivalent up to a minus. Here, we did not assume time-translation invariance at $t>s$. Therefore, taking the derivative with the respect to $t$ would not have brought out the operator $\lag^\dagger$ at time $s$ to the left of the propagation operator in the last term in Eq.~\eqref{eq:ABcorrOPERATORprep}. Rather, according to Eq.~\eqref{eq:FPAtdev}, it would have appeared to the right of the operator and would have been evaluated at time $t$; see Eq.~\eqref{eq:almost-HS} below. In time-invariant systems, $\lag^\dagger$ commutes with itself, hence the order of the adjoint propagator and $\lag^\dagger$ would not have mattered, and thus $\d/\d t$ would have given the desired result.

\subsection{The Harada-Sasa relation}

Particularly useful are correlations among positions, $\langle\vecX_t\vecX_s\rangle$. Not only are they (comparably-) easy to extract from experiments and simulations, they are connected to the response to a constant potential gradient (induced, \eg by gravity), $H(\vecx,s)\to H(\vecx,s)+\delta\vech_H(s)\cdot \vecx$ along the direction $\delta\vech_H/|\delta\vech_H|$. The equilibrium fluctuation-dissipation theorem for correlations among positions reads
\begin{equation}
    \frac{\d}{\d s}\langle X^\mu_t X^\nu_s\rangle=-\kt\frac{\delta\langle X_t^\mu\rangle}{\delta h_H^\nu(s)},\label{eq:FDTxx}
\end{equation}
where Greek indices denote different spatial directions.

Harada and Sasa~\cite{HaradaPRL05} investigated the breaking of the fluctuation-dissipation theorem in systems that are not in equilibrium, namely, those for which $\vecJ(\vecx,s)\neq\veco$. What they found is that the heat,
\begin{equation}
    \langle\dot Q_t\rangle=-\int\d\vecx\vecJ(\vecx,t)\cdot\vecF(\vecx,t),\label{eq:heat}
\end{equation}
which incorporates the nonzero flux $\vecJ(\vecx,s)$ in it, is exactly the difference between the correlation and the response functions,\begin{equation}
    \lim_{t\to s^+}\left[\frac{\d}{\d t}\frac{\d}{\d s}\langle \vecX_t\cdot\vecX_s\rangle+\kt\frac{\d}{\d t}\frac{\delta}{\delta \vech_H(s)}\cdot\langle \vecX_t\rangle\right]=-\mu(s)\langle\dot Q_s\rangle.\label{eq:HS}
\end{equation}
This can be seen by the fact that both the heat and the second term in Eq.~\eqref{eq:ABdev} that we have dropped for $\vecJ(\vecx,s)=\veco$ take the form $\int \d\vecx\vecJ(\vecx,s)[\cdot](\vecx,s)$. In addition to a valid Einstein relation, they had to assume a uniform scalar mobility, $\bmu=\mu(s)\vecI$. 

The usefulness of the Harada-Sasa relation~\cite{HaradaPRL05} is twofold. First, they found that this extra term is in fact a macroscopic and physically-meaningful observable. Second, it given an alternative route to measuring the heat instead of via calorimetry, requiring to observe the breaking of the fluctuation-dissipation theorem. This is particularly important in meso- and microscopic systems where the measurement of heat is not trivial. We now show how to derive the Harada-Sasa relation and the assumptions it requires.

First, we assume that the Einstein relation is valid at a time instant $s$, $\vecD(\vecx,s)=\kt(s)\bmu(\vecx,s)$. Upon reintroducing the missing term of Eq.~\eqref{eq:ABdev} to Eq.~\eqref{eq:FDTxx} and contracting the indices $\mu,\nu$ in the latter, under the response to a linear potential gradient considered above, we have
\begin{equation}
    \frac{\d}{\d s}\langle \vecX_t\cdot\vecX_s\rangle+\kt\frac{\delta}{\delta \vech_H(s)}\cdot\langle \vecX_t\rangle=\int\d\vecx \vecJ(\vecx,s)\cdot\exp\left[\int_s^t\d t'\lag^\dagger(\vecx,t')\right]\vecx.\label{eq:HSprep}
\end{equation}
Taking a time derivative with respect to $t$ using Eq.~\eqref{eq:FPAsdev} and applying $\lag^\dagger$ [Eq.~\eqref{eq:FPadjoint}] on $\vecx$ gives
\begin{subequations}
\begin{align}
    &\frac{\d}{\d t}\frac{\d}{\d s}\langle \vecX_t\cdot\vecX_s\rangle+\kt\frac{\d}{\d t}\frac{\delta}{\delta \vech_H(s)}\cdot\langle \vecX_t\rangle
    =\int\d\vecx \vecJ(\vecx,s)\cdot\exp\left[\int_s^t\d t'\lag^\dagger(\vecx,t')\right]\lag^\dagger(\vecx,t)\vecx\\
    &\qquad\qquad=\int\d\vecx \vecJ(\vecx,s)\cdot\exp\left[\int_s^t\d t'\lag^\dagger(\vecx,t')\right][\bmu(\vecx,s)\cdot\vecF(\vecx,s)+\grad\cdot\vecD(\vecx,s)].\label{eq:almost-HS}
\end{align}
\end{subequations}
We then take the limit where $t\to s^+$ and insert the Einstein relation,
\begin{equation}
    \lim_{t\to s^+}\left[\frac{\d}{\d t}\frac{\d}{\d s}\langle \vecX_t\cdot\vecX_s\rangle+\kt\frac{\d}{\d t}\frac{\delta}{\delta \vech_H(s)}\cdot\langle \vecX_t\rangle\right]
    =\int\d\vecx \vecJ(\vecx,s)\cdot\bmu(\vecx,s)\cdot\vecF(\vecx,s)+\int\d\vecx \vecJ(\vecx,s)\grad:\vecD(\vecx,s).\label{eq:almost-HS-lim}
\end{equation}
The right-hand side is not yet the heat of Eq.~\eqref{eq:heat}.
Finally, to get the Harada-Sasa relation, we must assume space-independent \emph{and} scalar mobility, $\bmu=\mu(s)\vecI$, \emph{in addition to the Einstein relation} [meaning that $\vecD=\kt(s)\mu(s)\vecI$ as well], which yields the Harada-Sasa relation, Eq.~\eqref{eq:HS}.

Indeed, the original Harada-Sasa relation~\cite{HaradaPRL05} and its various generalizations (\eg Refs.~\cite{fodor2016EP,NardiniPRX17,MartinPRE21}) all use constant and scalar mobility (and thus diffusivity, due to the Einstein relation) in their equations of motion. Assuming time-translation invariance and a valid Einstein relation, it is possible to reproduce the Harada-Sasa relation by considering the observable $(\d/\d t)\vecA(\vecX_t)=\bmu(\vecX_t)\dot\vecX_t$ instead of $(\d/\d t)\vecA(\vecX_t)=\dot\vecX_t$ for cases with a non-diagonal and configuration-dependent mobility~\cite{GolestanianARXIV2024}. Much like the observables we propose in Sec.~\ref{sec:new} for the case of violated Einstein relation, this observable may be nontrivial to estimate in experiment. 

To conclude, to get the fluctuation-dissipation theorem, we have assumed that the system was in equilibrium before the perturbation, meaning that the Einstein relations holds ($\vecD=\kt\bmu$) and that the system is flux-free ($\vecJ=\veco$). The Harada-Sasa relation relaxed the assumption of a flux-free state. But at the same time, one must have both the Einstein relation ($\vecD=\kt\bmu$) and a space-independent scalar diffusivity [$\bmu=\mu(s)\vecI$]. These are quite stringent assumptions that may considerably limit the range of applicability of the Harada-Sasa relation out of equilibrium.

\section{Proof of Eqs.~(20) and~(21) of the main text}\label{sec:new}

In this section, we aim to generalize the fluctuation-dissipation theorem and Harada-Sasa relation to cases in which Einstein relation is invalid, so there is no uniform scalar $\kt$ relating $\vecD$ to $\bmu$ (due to different anisotropy properties of these two tensors and/or their different dependence on space; see main text). This means that even within a flux-free steady state $\vecJ(\vecx,s)=\veco$, the response to a perturbed potential [Eq.~\eqref{eq:resPOTENT}] may not be related to the correlation with this potential [Eq.~\eqref{eq:ABdev}], and clearly the classical fluctuation-dissipation theorem does not hold. We show how even without the Einstein relation, an equation reminiscent of the Harada-Sasa relation can be obtained for both the heat [Eq.~\eqref{eq:heat}] and the informatic heat, 
\begin{equation}
    \langle\dot \Omega_t\rangle=-\int\d\vecx\vecJ(\vecx,t)\vecD^{-1}(\vecx,t)\cdot\bmu(\vecx,t)\cdot\vecF(\vecx,t).\label{eq:infoheat}
\end{equation}

%A valid question would be whom does the Harada-Sasa relation connects the breaking of fluctuation-dissipation theorem to: heat or informatic heat? As we saw in Eq.~\eqref{eq:almost-HS-lim}, we only get the heat if we are able to contract $\vecJ$ and $\vecF$ while $\bmu$ escapes the integral. Conversely, we may get the informatic heat if we are able to insert $\vecD^{-1}$ into the integral and in between $\vecJ$ and $\bmu\cdot\vecF$. Each one of these scenarios only occurs if each of them is constant an scalar. Thus, choosing $\vecA(\vecx,s)=\vecx$ and taking the additional derivative $\d/\d t$ does not yield the Harada-Sasa relation even with the Einstein relation.

\subsection{Extending the fluctuation-dissipation theorem}\label{sec:FDTwoE}

We seek to relate correlations to response functions in the absence of Einstein relation. In Sec.~\ref{sec:FDTwE} we considered a perturbation to the Hamiltonian, $H(\vecx,s)\to H(\vecx,s)+h_H(s)\pertH(\vecx)$. As a result, the forces are modified as $\vecF(\vecx,s)\to\vecF(\vecx,s)-\delta h_H(s)\grad H(\vecx)$, and thus the adjoint Fokker-Planck operator changes by $\delta\lag(\vecx,s)=\delta h_H[-\bmu(\vecx,s)\cdot\grad\pertH(\vecx)]$. In the latter expression, the Einstein relation allowed us to replace $\bmu(\vecx,s)$ with $\vecD(\vecx,s)$. Since we do not wish to assume the Einstein relation, we consider instead a response to a seemingly peculiar perturbation, $\vecF(\vecx,s)\to\vecF(\vecx,s)-h_\Phi(s)\bmu^{-1}(\vecx,s)\cdot\vecD(\vecx,s)\cdot\grad\pertPhi(\vecx)$, where $\pertPhi$ is some function. With this perturbation, $\vecD$ appears instead of $\bmu$ in the variation of the adjoint Fokker-Planck operator, $\delta\lag(\vecx,s)=\delta h_\Phi(s)[-\vecD(\vecx,s)\cdot\grad\pertPhi(\vecx)]$. The response function for such perturbation is
\begin{equation}
    \frac{\delta\langle A(\vecX_t,t)\rangle}{\delta h_\Phi(s)}=-\int\d\vecx p(\vecx,s)[\vecD(\vecx,s)\cdot\grad\pertPhi(\vecx)]\cdot\grad\exp\left[\int_{s}^{t}\d t'\lag^\dagger(\vecx,t')\right]A(\vecx,t).\label{eq:resNonPot}
\end{equation}
(The question whether such a perturbation is experimentally realizable is a nontrivial one.)

In order to obtain a generalized fluctuation-dissipation relation, we now assume that $\vecJ(\vecx,s)=\veco$. (when this is not the case, we get an extension of the Harada-Sasa relations; see Sec.~\ref{sec:HSgen}.) This is the assumption that makes the second term in Eq.~\eqref{eq:ABdev} disappear as in Sec.~\ref{sec:FDTwE}. In the absence of Einstein relation, these $\vecJ=\veco$ steady states are not equilibrium states, and they are the ones that lead to the ambiguous generalized temperature, Eq. (15) of the main text; see discussion in the last paragraph before the conclusion section in the main text. We give a simplistic example of such a state in Sec.~\ref{sec:steadystate} below. The flux-free steady-state constraint, $\vecJ=\veco$, implies $\vecF=-\bmu^{-1}\cdot\vecD\cdot\grad\Phi$, where $\Phi$ is a potential-like function with which $e^{-\Phi}$ is the steady-state distribution. With this in mind, we see that the perturbation proposed above is equivalent to the perturbation $\Phi(\vecx,s)\to \Phi(\vecx,s)+h_\Phi(s)\pertPhi(\vecx)$.

Thus, assuming that one begins from a flux-free steady-state [$\vecJ(\vecx,s)=\veco$] but without the Einstein relation, upon comparing Eq.~\eqref{eq:resNonPot} to Eq.~\eqref{eq:ABdev}, and choosing $B(\vecX_s,s)=\pertPhi(\vecX_s)$, we find
\begin{equation}
    \frac{\d}{\d s}\langle A(\vecX_t,t)\pertPhi(\vecX_s)\rangle=-\frac{\delta\langle A(\vecX_t,t)\rangle}{\delta h_\Phi(s)}.\label{eq:FDTnoneq}
\end{equation}
This expression first appeared in Ref.~\cite{AgarwalZph72}. As mentioned above, creating a perturbation to $\Phi$ may not be as experimentally realizable as a direct perturbation of the Hamiltonian. However, if one could apply a force $\vecF$ for which the quantity $\vecD^{-1}(\vecx,s)\cdot\bmu(\vecx,s)\cdot\vecF(\vecx,s)$ (without Einstein relation) is derived from a potential $\Phi$, one could possibly perturb that potential. Obviously, if the Einstein relation is obeyed, $\vecD(\vecx,s)=\kt(s)\bmu(\vecx,s)$, we get back the classical fluctuation-dissipation theorem, Eq.~\eqref{eq:FDT}.

\subsection{Extending the Harada-Sasa relation}\label{sec:HSgen}

We continue by extending the Harada-Sasa relation. To do so, we relax the assumption from Sec.~\ref{sec:FDTwoE} for which the system is initially in a flux-free steady-state, such that now $\vecJ(\vecx,s)\neq\veco$. As in Sec.~\ref{sec:FDTwoE}, we still choose $B(\vecX_s,s)=\pertPhi(\vecX_s)$ in Eq.~\eqref{eq:ABdev}, which now gives
\begin{equation}
    \frac{\d}{\d s}\langle A(\vecX_t,t)\pertPhi(\vecX_s)\rangle+\frac{\delta\langle A(\vecX_t,t)\rangle}{\delta h_\Phi(s)}=\int\d\vecx \vecJ(\vecx,s)\cdot[\grad \pertPhi (\vecx,s)]\exp\left[\int_s^t\d t'\lag^\dagger(\vecx,t')\right]A(\vecx,t).\label{eq:genStartPt}
\end{equation}
As we have seen in Eq.~\eqref{eq:almost-HS-lim}, we cannot obtain the force through a derivative with respect to $t$. Thus, instead of choosing $\vecA(\vecx,t)=\vecx$, we choose the force as our observable, $\vecA(\vecx,t)=\vecF(\vecx,t)$. At the same time, we choose a linear $\pertPhi(\vecx)=\vecx$, and take the limit where $t$ is a time instant shortly after $s$,
\begin{equation}
    \lim_{t\to s^+}\left[\frac{\d}{\d s}\langle\vecX_s\cdot\vecF(\vecX_t,t)\rangle+\frac{\delta}{\delta \vech_\Phi(s)}\cdot\langle \vecF(\vecX_t,t)\rangle\right]%=\int\d\vecx \vecJ(\vecx,s)\cdot\vecF(\vecx,t)
    =-\langle\dot Q_s\rangle.\label{eq:genHS}
\end{equation}
This is Eq.~(20) of the main text. Thus, without the Einstein relation, also the heat is related to a peculiar response in addition to the cumbersome correlation.
On the other hand, by choosing an even less trivial observable, $\vecA(\vecx,s)=\vecD^{-1}(\vecx,t)\cdot\bmu(\vecx,t)\cdot\vecF(\vecx,t)$, we find a similar relation for the informatic heat,
\begin{equation}
    \lim_{t\to s^+}\left[\frac{\d}{\d s}\langle\vecX_s\cdot\vecD^{-1}(\vecX_t,t)\cdot\bmu(\vecX_t,t)\cdot\vecF(\vecX_t,t)\rangle+\frac{\delta}{\delta \vech_\Phi(s)}\cdot\langle \vecD^{-1}(\vecX_t,t)\cdot\bmu(\vecX_t,t)\cdot\vecF(\vecX_t,t)\rangle\right]%\\=\int\d\vecx \vecJ(\vecx,s)\cdot\vecD^{-1}(\vecx,t)\cdot\bmu(\vecx,t)\cdot\vecF(\vecx,t)
    =\langle\dot \Omega_s\rangle.\label{eq:infoHS}
\end{equation}
This is Eq.~(21) of the main text.

With the Einstein relation, the response is to a conservative force as before, and there is no distinction between the heat $\dot Q_s$ and the informatic heat $\dot\Omega_s$ up to the temperature. In order to get the usual Harada-Sasa relation~\cite{HaradaPRL05}, one should assume in addition that the mobility is uniform and scalar, $\bmu=\mu(s)\vecI$. This can be seen by multiplying both sides of Eq.~\eqref{eq:genHS} by $\mu$ [or, conversely, Eq.~\eqref{eq:infoHS} by $\kt\mu$] and insert $\mu$ into the ensemble averages on the left-hand side. Identifying that $\mu\vecF(\vecX_t,t)$ (note that $\grad\cdot\vecD$ with $\vecD=\kt\mu\vecI$ is zero) is the average of the velocity $(\d/\d t)\vecX_t$, we converge back to the usual Harada-Sasa relation, Eq.~\eqref{eq:HS}.

\section{Classical-thermodynamic relations}\label{sec:thermo}

In this section, we provide further details on the proof and the assumptions of the thermodynamic theorems that result from the extended second law of thermodynamics constructed in the main text,
\begin{equation}
\langle\dot\diss_t\rangle=\kb \teff\langle\dot S_t\rangle-\langle\dot Q_t\rangle\geq0.\label{eq:2nd}
\end{equation}
Reintroducing the protocol $\Lambda_t$, we recall the expressions described in the main text for the average heat, change in entropy, and heat dissipation:
\begin{eqnarray}
    \langle\dot Q_t\rangle&=&-\int \d\vecx\vecJ(\vecx,t)\cdot\vecF(\vecx,\Lambda_t)
    \nonumber\\&=&-\int \d\vecx p(\vecx,t)\left\{\vecF(\vecx,\Lambda_t)\cdot\bmu(\vecx,\Lambda_t)\cdot\vecF(\vecx,\Lambda_t)+ \grad\cdot\left[\vecD(\vecx,\Lambda_t)\cdot\vecF(\vecx,\Lambda_t)\right]\right\},\label{eq:Qdot}
    \\
    \langle\dot S_t\rangle&=&-\int \d\vecx\vecJ(\vecx,t)\cdot\grad\ln p(\vecx,t)=\int \d\vecx[\grad\cdot\vecJ(\vecx,t)]\ln p(\vecx,t)
    \nonumber\\&=&-\int \d\vecx p(\vecx,t)\left\{\vecF(\vecx,\Lambda_t)\cdot\bmu(\vecx,\Lambda_t)\cdot\frac{\grad p(\vecx,t)}{p(\vecx,t)}+\grad\cdot\left[\vecD(\vecx,\Lambda_t)\cdot\frac{\grad p(\vecx,t)}{p(\vecx,t)}\right]\right\},\label{eq:Sdot}
    \\
    \langle\dot \diss_t\rangle&=&\kb \teff\int \d\vecx \frac{\vecJ(\vecx,t)\cdot\vecD^{-1}(\vecx,\Lambda_t)\cdot\vecJ(\vecx,t)}{p(\vecx,t)}.\label{eq:DISSdot}
\end{eqnarray}
The nonequilibrium temperature that we identified in the main text is
\begin{eqnarray}
    \frac{1}{\kb \teff(t)}&=&\frac{\int \d\vecx\vecJ(\vecx,t)\cdot\vecD^{-1}(\vecx,\Lambda_t)\cdot[\bmu(\vecx,\Lambda_t)\cdot\vecF(\vecx,\Lambda_t)]}{\int \d\vecx\vecJ(\vecx,t)\cdot\vecF(\vecx,\Lambda_t)}\nonumber\\
    &=&\frac{\int \d\vecx p(\vecx,t)\left\{[\bmu(\vecx,\Lambda_t)\cdot\vecF(\vecx,\Lambda_t)]\cdot\vecD^{-1}(\vecx,\Lambda_t)\cdot[\bmu(\vecx,\Lambda_t)\cdot\vecF(\vecx,\Lambda_t)]+ \grad\cdot\left[\bmu(\vecx,\Lambda_t)\cdot\vecF(\vecx,\Lambda_t)\right]\right\}}{\int \d\vecx p(\vecx,t)\left\{\vecF(\vecx,\Lambda_t)\cdot\bmu(\vecx,\Lambda_t)\cdot\vecF(\vecx,\Lambda_t)+ \grad\cdot\left[\vecD(\vecx,\Lambda_t)\cdot\vecF(\vecx,\Lambda_t)\right]\right\}}.\label{eq:temp}
\end{eqnarray}
It depends on both the protocol (through the variation of the forcing $\vecF$, mobility $\bmu$, and diffusivity $\vecD$) and on time (through the dynamics that the distribution $p$ undergoes given the protocol evolution). We identify the connection between informatic entropy production and heat dissipation, $\langle\dot\Sigma_t\rangle=\langle\dot\diss_t\rangle/(\kb \teff)$, and between the informatic and calorimetric heats, $\langle\dot\Omega_t\rangle=\langle\dot Q_t\rangle/(\kb \teff)$.

\subsection{Clausius inequality}

Within classical thermodynamics, the Clausius inequality states that for any transformation (reversible or not) starting and ending at the same equilibrium state, $\oint \d Q/T_\mathrm{env}\leq0$, where $\d Q$ are the heat increments, $T_\mathrm{env}$ is the temperature of the environment, and the closed integral is parameterized with the protocol value. We now investigate the fate of this theorem with our generalized temperature.

Consider a process under a circular protocol $\Lambda_t=\lambda^\circ(t)$, starting at steady state. Quantities with the circle superscript denote that they are computed for this circular protocol. Between times $t\in[0,t^\circ]$, the protocol changes such that $\lambda^\circ(0)=\lambda^\circ(t^\circ)\equiv\lambda^\circ_0$. Once it reaches that value, it remains constant, $\lambda^\circ(t>t^\circ)=\lambda^\circ_0$, so the system may now decay under the constant protocol value to its corresponding steady-state distribution. We assume that the system has a finite relaxation time for that protocol value, which we denote $\tau(\lambda^\circ_0)$. Thus, when the protocol reaches $\lambda^\circ_0$ at time $t^\circ$, the system reaches the steady-state distribution (the same as the initial one) only after an additional transient, at time $t=\tf^\circ\gg t^\circ+\tau(\lambda^\circ_0)$, meaning that $p(\vecx,\tf^\circ)\to p(\vecx,0)$. 

Since the system begins at and approaches to the same state, the entropy\,---\,a state function\,---\,does not change during this process, 
\begin{equation}
\langle S^\circ_{\tf^\circ}\rangle-\langle S^\circ_0\rangle=\int_0^{\tf^\circ} \d t\langle\dot{S}^\circ_t\rangle=0.\label{eq:clausiusentropy}
\end{equation}
At every instant during this process, the generalized temperature $\teff^\circ(t)$ is given by Eq.~\eqref{eq:temp}, the average heat rate $\langle \dot{Q}_t^\circ\rangle$ by Eq.~\eqref{eq:Qdot}, and the average heat dissipation rate $\langle \dot{\diss}_t^\circ\rangle=\kb\teff^\circ(t)\langle \dot{\Sigma}_t^\circ\rangle$ by Eq.~\eqref{eq:DISSdot}. Combining that with Eq.~\eqref{eq:2nd}, we find the nonequilibrium analogue to the Clausius inequality,
\begin{equation}
    \int_0^{\tf^\circ} \d t\frac{\langle\dot{Q}^\circ_t\rangle}{\kb \teff^\circ(t)}=-\int_0^{\tf^\circ} \d t\frac{\langle\dot{\diss}^\circ_t\rangle}{\kb \teff^\circ(t)}=-\int_0^{\tf^\circ} \d t\langle\dot{\Sigma}^\circ_t\rangle\leq0.\label{eq:clausius}
\end{equation}
[The integrals in Eqs.~\eqref{eq:clausiusentropy} and~\eqref{eq:clausius} are between $0$ and $\tf^\circ$ because only at $\tf^\circ\gg t^\circ+\tau(\lambda^\circ_0)$ the system reaches a steady-state.]

Comparing Eq.~\eqref{eq:clausius} and the classical Clausius inequality, we make a few remarks: 
\begin{itemize}
    \item The temperature in the Clausius inequality is of the heat reservoir (which is of infinite size and in equilibrium). In our case, other than setting $\vecD(\vecx,\Lambda_t)$ and $\bmu(\vecx,\Lambda_t)$, the medium in which the system is immersed does not have any specified properties. Therefore, we perceive the nonequilibrium temperature $\teff$ as the ``system's'' temperature. At best, inspired by the zeroth law, we may set the medium's temperature, for constant $\Lambda_t=\lambda$, as the limiting temperature of the system in contact, $\teff(t\gg\tau(\lambda))$.
    \item For reversible (near-equilibrium) processes, the Clausius inequality is an equality. Here, even if the Einstein relation does not apply, we find that at flux-free steady state, $\vecJ=\veco$, the generalized Clausius ineqsuality becomes an equality. Moreover, it indicates that the flux-free state is stable (that is, for a small departure from equilibrium $\delta\vecJ$, the dissipation is quadratic in it and hence much smaller). Lastly, based on various coarse graining techniques of the informatic entropy production (e.g., the thermodynamic uncertainty relations~\cite{GingrichPRL16,SeifertPRL15}), if a positive estimate of $\langle\Sigma_t\rangle$ is found,one obtains a tighter bound than the usual Clausius inequality.
\end{itemize}

\subsection{Efficiency of a heat engine}

In this section, we consider the extension of the Carnot efficiency theorem with our generalized temperature. To this end, we consider $\Lambda_t$ that is a combination of at least two protocols. Notice that the nonequilibrium temperature $\teff(t)$ and average heat rate $\langle\dot Q_t\rangle$ are determined solely by the imposed protocol $\Lambda_t$ and the initial conditions $p(\vecx,0)$. This can be seen by inserting $p(\vecx,t)=\exp[\int_0^t\hat{L}(\vecx,\Lambda_t)]p(\vecx,0)$ in Eqs.~\eqref{eq:temp} and~\eqref{eq:Qdot}. For a set of initial condition $p(\vecx,0)$, we realize that the condition $\langle\dot Q_t\rangle=0$ defines an adiabatic manifold in protocol space (e.g., if the forcing is such that  $\vecF(\vecx,\Lambda_t)\cdot\bmu(\vecx,\Lambda_t)\cdot\vecF(\vecx,\Lambda_t)=-\grad\cdot\left[\vecD(\vecx,\Lambda_t)\cdot\vecF(\vecx,\Lambda_t)\right]$, or if the system's container is isolated), and the condition $\partial\teff(t)/\partial t=0$ gives a first-order differential equation for $\Lambda_t$ (under the initial condition of the desired temperature $\teff$ for $\Lambda_0$) which determines an ``isothermal'' surface. 
We comment that, under the nonequilibrium setup of the main text, it may be nontrivial to realize such ``isothermal'' process. This is why extending the zeroth law of thermodynamics could be quite beneficial, as then an appropriate (nonequilibrium) environment would enforce a constant temperature, just as having an isolated container ensures an adiabatic process. 

We construct the following cycle out of the initial condition $p(\vecx,0)$:
\begin{enumerate}
    \item An isothermal protocol $\Lambda_t=\lambda^{\teff}_\mathrm{H}(t)$ acting during $t\in[0,t^{\teff}_\mathrm{H}]$ for which $\teff(t)=\teff_\mathrm{H}=\mathrm{const}$. The average accumulated heat in this step is 
    \begin{equation}
        \langle Q_\mathrm{in}\rangle\equiv\int_0^{t^{\teff}_\mathrm{H}}\d t \langle\dot Q_t\rangle.
    \end{equation}
    The protocol $\lambda^{\teff}_\mathrm{H}(t)$ is such that $\langle Q_\mathrm{in}\rangle>0$.
    \item An adiabatic transformation $\Lambda_t=\lambda^Q_{\mathrm{H}\to \mathrm{C}}(t)$ during $t\in[t^{\teff}_\mathrm{H},t^{\teff}_\mathrm{H}+t^Q_{\mathrm{H}\to \mathrm{C}}]$ that reaches a state $p(\vecx,t^{\teff}_\mathrm{H}+t^Q_{\mathrm{H}\to \mathrm{C}})$ for which we have $\teff(t^{\teff}_\mathrm{H}+t^Q_{\mathrm{H}\to \mathrm{C}})=\teff_\mathrm{C}$. As an adiabatic process, $\langle Q_t\rangle=0$ during this step.
    \item Another isothermal protocol $\Lambda_t=\lambda^{\teff}_\mathrm{C}(t)$ acting during $t\in[t^{\teff}_\mathrm{H}+t^Q_{\mathrm{H}\to \mathrm{C}},t^{\teff}_\mathrm{H}+t^Q_{\mathrm{H}\to \mathrm{C}}+t^{\teff}_\mathrm{C}]$ for which $\teff(t)=\teff_\mathrm{C}=\mathrm{const}$. The average accumulated heat in this step is 
    \begin{equation}
        \langle Q_\mathrm{out}\rangle\equiv\int_{t^{\teff}_\mathrm{H}+t^Q_{\mathrm{H}\to \mathrm{C}}}^{t^{\teff}_\mathrm{H}+t^Q_{\mathrm{H}\to \mathrm{C}}+t^{\teff}_\mathrm{C}}\d t \langle\dot Q_t\rangle,
    \end{equation}
    In order for the engine to be at all useful, the protocol should be tailored such that $\langle Q_\mathrm{out}\rangle<0$.
    \item Another adiabatic transformation $\Lambda_t=\lambda^Q_{\mathrm{C}\to \mathrm{H}}(t)$ during $t\in[t^{\teff}_\mathrm{H}+t^Q_{\mathrm{H}\to \mathrm{C}}+t^{\teff}_\mathrm{C},\tf]$, where $\tf=t^{\teff}_\mathrm{H}+t^Q_{\mathrm{H}\to \mathrm{C}}+t^{\teff}_\mathrm{C}+t^Q_{\mathrm{C}\to \mathrm{H}}$, that reaches exactly the initial $p(\vecx,\tf)=p(\vecx,0)$ with $\lambda^Q_{\mathrm{C}\to \mathrm{H}}(\tf)=\lambda^{\teff}_\mathrm{H}(0)$, such that indeed $\teff(\tf)=\teff_\mathrm{H}$. As an adiabatic process, $\langle Q_t\rangle=0$ during this step as well.
\end{enumerate}
The total informatic entropy production over the four steps is denoted as
\begin{equation}
    \langle\Sigma_\mathrm{cycle}\rangle\equiv\int_{0}^{t^{\teff}_\mathrm{H}}\d t\langle\dot\Sigma_t\rangle+\int_{t^{\teff}_\mathrm{H}}^{t^{\teff}_\mathrm{H}+t^Q_{\mathrm{H}\to \mathrm{C}}}\d t\langle\dot\Sigma_t\rangle+\int_{t^{\teff}_\mathrm{H}+t^Q_{\mathrm{H}\to \mathrm{C}}}^{t^{\teff}_\mathrm{H}+t^Q_{\mathrm{H}\to \mathrm{C}}+t^{\teff}_\mathrm{C}}\d t\langle\dot\Sigma_t\rangle+\int_{t^{\teff}_\mathrm{H}+t^Q_{\mathrm{H}\to \mathrm{C}}+t^{\teff}_\mathrm{C}}^{\tf}\d t\langle\dot\Sigma_t\rangle
    \\=\int_0^{\tf}\d t\langle\dot\Sigma_t\rangle\geq0.
\end{equation}

We proceed to compute the engine's efficiency, $\langle-W_\mathrm{cycle}\rangle/\langle Q_\mathrm{in}\rangle$, where $\langle-W_\mathrm{cycle}\rangle\equiv-\int_0^{\tf}\d t\langle\dot W_t\rangle$ is the total work extracted from the system during one cycle.
Since we returned to $p(\vecx,\tf)=p(\vecx,0)$ and $\lambda^Q_{\mathrm{C}\to \mathrm{H}}(\tf)=\lambda^{\teff}_\mathrm{H}(0)$, the mean energy has not changed during one cycle, $\int_0^{\tf}\d t\langle\dot H_t\rangle=0$. Therefore, according to the first law of thermodynamics~\cite{sekimoto98Langevin_thermo,book:sekimoto} [Eq. (7) of the main text], the extracted work is $\langle-W_\mathrm{cycle}\rangle=\langle Q_\mathrm{in}\rangle+\langle Q_\mathrm{out}\rangle$. 
Now, since $\teff$ is constant for the first and third steps and no heat is involved in the second and fourth steps, and according to Eq.~\eqref{eq:clausius} for one cycle we have
\begin{equation}
    \frac{\langle Q_\mathrm{in}\rangle}{\kb \teff_\mathrm{H}}+\frac{\langle Q_\mathrm{out}\rangle}{\kb \teff_\mathrm{C}}=-\int_0^{\tf}\d t\frac{\langle\dot\diss_t\rangle}{\kb \teff(t)} =-\langle\Sigma_\mathrm{cycle}\rangle.
\end{equation}
Thus, we arrive at 
\begin{equation}
    \frac{\langle-W_\mathrm{cycle}\rangle}{\langle Q_\mathrm{in}\rangle}=1+\frac{\langle Q_\mathrm{out}\rangle}{\langle Q_\mathrm{in}\rangle}=1-\frac{{\teff}_\mathrm{C}}{{\teff}_\mathrm{H}}-\frac{\kb \teff_\mathrm{C}\langle \Sigma_\mathrm{cycle}\rangle}{\langle Q_\mathrm{in}\rangle}.\label{eq:carnot}
\end{equation}
Unless one has an estimate of the heat dissipation (informatic entropy production) along a cycle, inserting the bound $\langle \Sigma_\mathrm{cycle}\rangle\geq0$ makes this expression coincide with the Carnot bound for the efficiency of a heat engine, $\langle-W_\mathrm{cycle}\rangle/\langle Q_\mathrm{in}\rangle\leq1-{\teff}_\mathrm{C}/{\teff}_\mathrm{H}$. If the processes were near equilibrium throughout, then $\langle \Sigma_\mathrm{cycle}\rangle=0$, and we exactly reach Carnot efficiency.

Similar to the discussion following our version of the Clausius inequality, it remains to be explored if the temperatures in Eq.~\eqref{eq:carnot} are more than just obscure integrated properties of the system which, through a meticulous choice of protocols, were kept constant. After all, the temperatures that appear in the classical Carnot bound are those of the two baths, such that the temperature of the system must not coincide with the ones of the baths during irreversible processes.

\subsection{Changes in free energy}

It is well known that nonequilibrium steady-states do not necessarily obey a minimum free energy principle~\cite{TakatoriPRE15,MartinPRE21}. However, for isothermal transformations we are able to define a free energy that leads to two more relations. We begin, once again, from some $p(\vecx,0)$ and perform a protocol $\Lambda(t)=\lambda^{\teff}(t)$ which maintains a constant temperature $\teff$.

Define the nonequilibrium Helmholtz free energy $A_t=H_t-\kb \teff S_t$. Assuming the temperature is indeed constant, $\dot A_t=\dot H_t-\kb\teff \dot S_t$. Upon inserting Eq.~\eqref{eq:2nd} and identifying the work $\dot W_t=\dot H_t-\dot Q_t$, we find
\begin{equation}
    \dot W_t=\dot A_t+\dot\diss_t.
\end{equation}
Taking the ensemble average, we find a bound on the extractable work in noncyclic 
isothermal processes,
\begin{equation}
    \langle-\dot W_t\rangle=-(\langle\dot A_t\rangle+\langle\dot\diss_t\rangle)\leq-\langle\dot A_t\rangle,\label{eq:WdAineq}
\end{equation}
which is analogous to the classical relation between extractable work and the change of free energy. As mentioned in the main text, since the informatic heat and calorimetric heat are only equal on average and differ otherwise, the Jarzynski equality~\cite{JarzynskiPRL97,JarzynskiPRE97} (that could have reshaped Eq.~\eqref{eq:WdAineq} into an equality) is invalid.

Lastly, we investigate isothermal processes which occur without external interventions, that is, $\dot W_t=0$. (This means that $\vecF=-\grad H$.) We arrive at
\begin{equation}
    \langle\dot A_t\rangle=-\langle\dot\diss_t\rangle\leq0.\label{eq:Avar}
\end{equation}
This means that the generalized free energy, like its classical counterpart, keeps decreasing until it reaches steady state in spontaneous isothermal processes. This proposes a variational principle for steady-states that result from spontaneous isothermal evolution. 

\section{Example: Steady states with ambiguous generalized temperature}\label{sec:steadystate}

In this section we discuss a peculiarity related to the nonequilibrium temperature introduced in the main text. With a valid Einstein relation $\bmu=\vecD/(\kt)$ and time-independent conservative forces $\vecF(\vecx)=-\grad H(\vecx)$, the steady-state distribution of the Fokker-Planck equation
\begin{equation}
    \frac{\partial p(\vecx,t)}{\partial t}=-\grad\cdot\vecJ(\vecx,t),\qquad\vecJ(\vecx,t)=-\vecD(\vecx,t)\cdot\left[\frac{\grad H(\vecx)}{\kt}p(\vecx,t)+\grad p(\vecx,t)\right]
\end{equation}
is clearly the Boltzmann distribution $p(\vecx,t\to\infty)\sim e^{-H(\vecx)/(\kt)}$ with the Einstein $T$.
Now consider the peculiar forcing considered in Sec.~\ref{sec:FDTwoE}, where $\vecF(\vecx,t)=-\bmu^{-1}(\vecx,t)\cdot\vecD(\vecx,t)\cdot\grad\Phi(\vecx)$ with a scalar function $\Phi(\vecx)$. The resulting Fokker-Planck equation,
\begin{equation}
    \frac{\partial p(\vecx,t)}{\partial t}=-\grad\cdot\vecJ(\vecx,t),\qquad\vecJ(\vecx,t)=-\vecD(\vecx,t)\cdot\left[p(\vecx,t)\grad \Phi(\vecx)+\grad p(\vecx,t)\right],
\end{equation}
also has a Boltzmann-like steady-state, $p(\vecx,t\to\infty)\sim e^{-\Phi(\vecx)}$, although no temperature appears in the exponent. Moreover, $\vecJ(\vecx,t\to\infty)\to\veco$, meaning that the proposed nonequilibrium temperature [Eq. (15) of the main text] has a `$0/0$' ambiguity.

In an attempt to mend this peculiarity, we consider a two-dimensional system approaching this flux-free steady state. Assume that it has constant and diagonal $\bmu=\mathrm{diag}(\mu_1,\mu_2)$ and $\vecD=\mathrm{diag}(D_1=\kt_1\mu_1,D_2=\kt_2\mu_2)$. For this simplified model of two ``particles'', an effective temperature was identified in Ref.~\cite{JoannyPRE15} for the separation vector among them. The nonequilibrium temperature we propose here has the following weighted average, 
\begin{equation}
    \frac{1}{\kb \teff(t)}=\frac{[\int d\vecx J_1(\vecx,t)F_1(\vecx,\lambda(t))]/(\kt_1)+[\int d\vecx J_2(\vecx,t)F_2(\vecx,\lambda(t))]/(\kt_2)}{[\int d\vecx J_1(\vecx,t)F_1(\vecx,\lambda(t))]+[\int d\vecx J_2(\vecx,t)F_2(\vecx,\lambda(t))]}, 
\end{equation}
Upon approach to steady state, the limit $\vecJ\to\veco$ is, of course, ill-defined, and it depends on the interplay between the heat rates along each axis, $[\int d\vecx J_1(\vecx,t)F_1(\vecx,\lambda(t))]/[\int d\vecx J_2(\vecx,t)F_2(\vecx,\lambda(t))]$. For example, the effective temperature for the separation vector of Ref.~\cite{JoannyPRE15}, $1/(\kb\teff)=[(1/\mu_1)+(1/\mu_1)]/[(\kt_1/\mu_1)+(\kt_2/\mu_1)]$, can readily be obtained by justifying the ratio among dissipations $[\int d\vecx J_1(\vecx,t)F_1(\vecx,\lambda(t))]/[\int d\vecx J_2(\vecx,t)F_2(\vecx,\lambda(t))]=(\kt_1/\mu_1)/(\kt_2/\mu_2)$. In other words, depending on what variable has been perturbed away from steady state (e.g., the separation vector as in Ref.~\cite{JoannyPRE15}), $(\kb\teff)^{-1}$ will be given by a different linear combination of $1/(\kt_1)$ and $1/(\kt_2)$. In some sense, this dependence on the axis of relaxation isolates the dominant contribution in terms of dissipation~\cite{ArielJSP15,ArielPHYSA19}.
%Upon solving (`diagonalizing') the Fokker-Planck equation [Eq.~\eqref{eq:FPE}] with many dimensional, nondiagonal, space- and protocol-dependent $\bmu$ and $\vecD$, identical behaviour will be seen. We hope that the thermodynamic properties of systems violating the ER, for which the object $\vecD^{-1}\cdot(\bmu\cdot\vecF)$ derives from a potential $\Phi$ will be further studied.

\end{widetext}


\begin{thebibliography}{90}%
\makeatletter
\providecommand \@ifxundefined [1]{%
 \@ifx{#1\undefined}
}%
\providecommand \@ifnum [1]{%
 \ifnum #1\expandafter \@firstoftwo
 \else \expandafter \@secondoftwo
 \fi
}%
\providecommand \@ifx [1]{%
 \ifx #1\expandafter \@firstoftwo
 \else \expandafter \@secondoftwo
 \fi
}%
\providecommand \natexlab [1]{#1}%
\providecommand \enquote  [1]{``#1''}%
\providecommand \bibnamefont  [1]{#1}%
\providecommand \bibfnamefont [1]{#1}%
\providecommand \citenamefont [1]{#1}%
\providecommand \href@noop [0]{\@secondoftwo}%
\providecommand \href [0]{\begingroup \@sanitize@url \@href}%
\providecommand \@href[1]{\@@startlink{#1}\@@href}%
\providecommand \@@href[1]{\endgroup#1\@@endlink}%
\providecommand \@sanitize@url [0]{\catcode `\\12\catcode `\$12\catcode
  `\&12\catcode `\#12\catcode `\^12\catcode `\_12\catcode `\%12\relax}%
\providecommand \@@startlink[1]{}%
\providecommand \@@endlink[0]{}%
\providecommand \url  [0]{\begingroup\@sanitize@url \@url }%
\providecommand \@url [1]{\endgroup\@href {#1}{\urlprefix }}%
\providecommand \urlprefix  [0]{URL }%
\providecommand \Eprint [0]{\href }%
\providecommand \doibase [0]{https://doi.org/}%
\providecommand \selectlanguage [0]{\@gobble}%
\providecommand \bibinfo  [0]{\@secondoftwo}%
\providecommand \bibfield  [0]{\@secondoftwo}%
\providecommand \translation [1]{[#1]}%
\providecommand \BibitemOpen [0]{}%
\providecommand \bibitemStop [0]{}%
\providecommand \bibitemNoStop [0]{.\EOS\space}%
\providecommand \EOS [0]{\spacefactor3000\relax}%
\providecommand \BibitemShut  [1]{\csname bibitem#1\endcsname}%
\let\auto@bib@innerbib\@empty
%</preamble>
\bibitem [{\citenamefont {Ramaswamy}(2010)}]{Rev:Ram10}%
  \BibitemOpen
  \bibfield  {author} {\bibinfo {author} {\bibfnamefont {S.}~\bibnamefont
  {Ramaswamy}},\ }\bibfield  {title} {\bibinfo {title} {The mechanics and
  statistics of active matter},\ }\href
  {https://doi.org/10.1146/annurev-conmatphys-070909-104101} {\bibfield
  {journal} {\bibinfo  {journal} {Annu. Rev. Condens. Matter Phys.}\ }\textbf
  {\bibinfo {volume} {1}},\ \bibinfo {pages} {323} (\bibinfo {year}
  {2010})}\BibitemShut {NoStop}%
\bibitem [{\citenamefont {Marchetti}\ \emph {et~al.}(2013)\citenamefont
  {Marchetti}, \citenamefont {Joanny}, \citenamefont {Ramaswamy}, \citenamefont
  {Liverpool}, \citenamefont {Prost}, \citenamefont {Rao},\ and\ \citenamefont
  {Simha}}]{Rev:MarRam12}%
  \BibitemOpen
  \bibfield  {author} {\bibinfo {author} {\bibfnamefont {M.~C.}\ \bibnamefont
  {Marchetti}}, \bibinfo {author} {\bibfnamefont {J.-F.}\ \bibnamefont
  {Joanny}}, \bibinfo {author} {\bibfnamefont {S.}~\bibnamefont {Ramaswamy}},
  \bibinfo {author} {\bibfnamefont {T.~B.}\ \bibnamefont {Liverpool}}, \bibinfo
  {author} {\bibfnamefont {J.}~\bibnamefont {Prost}}, \bibinfo {author}
  {\bibfnamefont {M.}~\bibnamefont {Rao}},\ and\ \bibinfo {author}
  {\bibfnamefont {R.~A.}\ \bibnamefont {Simha}},\ }\bibfield  {title} {\bibinfo
  {title} {Hydrodynamics of soft active matter},\ }\href
  {https://doi.org/10.1103/RevModPhys.85.1143} {\bibfield  {journal} {\bibinfo
  {journal} {Rev. Mod. Phys.}\ }\textbf {\bibinfo {volume} {85}},\ \bibinfo
  {pages} {1143} (\bibinfo {year} {2013})}\BibitemShut {NoStop}%
\bibitem [{\citenamefont {Be'er}\ and\ \citenamefont
  {Ariel}(2019)}]{REV:bacteria}%
  \BibitemOpen
  \bibfield  {author} {\bibinfo {author} {\bibfnamefont {A.}~\bibnamefont
  {Be'er}}\ and\ \bibinfo {author} {\bibfnamefont {G.}~\bibnamefont {Ariel}},\
  }\bibfield  {title} {\bibinfo {title} {A statistical physics view of swarming
  bacteria},\ }\href {https://doi.org/10.1186/s40462-019-0153-9} {\bibfield
  {journal} {\bibinfo  {journal} {Mov. Ecol.}\ }\textbf {\bibinfo {volume}
  {7}},\ \bibinfo {pages} {9} (\bibinfo {year} {2019})}\BibitemShut {NoStop}%
\bibitem [{\citenamefont {Schmiedl}\ and\ \citenamefont
  {Seifert}(2007)}]{SiefertJCP07}%
  \BibitemOpen
  \bibfield  {author} {\bibinfo {author} {\bibfnamefont {T.}~\bibnamefont
  {Schmiedl}}\ and\ \bibinfo {author} {\bibfnamefont {U.}~\bibnamefont
  {Seifert}},\ }\bibfield  {title} {\bibinfo {title} {Stochastic thermodynamics
  of chemical reaction networks},\ }\href {https://doi.org/10.1063/1.2428297}
  {\bibfield  {journal} {\bibinfo  {journal} {J. Chem. Phys.}\ }\textbf
  {\bibinfo {volume} {126}},\ \bibinfo {pages} {044101} (\bibinfo {year}
  {2007})}\BibitemShut {NoStop}%
\bibitem [{\citenamefont {Rao}\ and\ \citenamefont
  {Esposito}(2018)}]{RaoJCP18}%
  \BibitemOpen
  \bibfield  {author} {\bibinfo {author} {\bibfnamefont {R.}~\bibnamefont
  {Rao}}\ and\ \bibinfo {author} {\bibfnamefont {M.}~\bibnamefont {Esposito}},\
  }\bibfield  {title} {\bibinfo {title} {{Conservation laws and work
  fluctuation relations in chemical reaction networks}},\ }\href
  {https://doi.org/10.1063/1.5042253} {\bibfield  {journal} {\bibinfo
  {journal} {J. Chem. Phys.}\ }\textbf {\bibinfo {volume} {149}},\ \bibinfo
  {pages} {245101} (\bibinfo {year} {2018})}\BibitemShut {NoStop}%
\bibitem [{\citenamefont {Bernheim-Groswasser}\ \emph
  {et~al.}(2018)\citenamefont {Bernheim-Groswasser}, \citenamefont {Gov},
  \citenamefont {Safran},\ and\ \citenamefont {Tzlil}}]{GovRev18}%
  \BibitemOpen
  \bibfield  {author} {\bibinfo {author} {\bibfnamefont {A.}~\bibnamefont
  {Bernheim-Groswasser}}, \bibinfo {author} {\bibfnamefont {N.~S.}\
  \bibnamefont {Gov}}, \bibinfo {author} {\bibfnamefont {S.~A.}\ \bibnamefont
  {Safran}},\ and\ \bibinfo {author} {\bibfnamefont {S.}~\bibnamefont
  {Tzlil}},\ }\bibfield  {title} {\bibinfo {title} {Living matter: Mesoscopic
  active materials},\ }\href
  {https://doi.org/https://doi.org/10.1002/adma.201707028} {\bibfield
  {journal} {\bibinfo  {journal} {Adv. Mater.}\ }\textbf {\bibinfo {volume}
  {30}},\ \bibinfo {pages} {1707028} (\bibinfo {year} {2018})}\BibitemShut
  {NoStop}%
\bibitem [{\citenamefont {Casas-Vazquez}\ and\ \citenamefont
  {Jou}(2003)}]{CasasVazquez2003}%
  \BibitemOpen
  \bibfield  {author} {\bibinfo {author} {\bibfnamefont {J.}~\bibnamefont
  {Casas-Vazquez}}\ and\ \bibinfo {author} {\bibfnamefont {D.}~\bibnamefont
  {Jou}},\ }\bibfield  {title} {\bibinfo {title} {{Temperature in
  non-equilibrium states: A review of open problems and current proposals}},\
  }\href {https://doi.org/10.1088/0034-4885/66/11/R03} {\bibfield  {journal}
  {\bibinfo  {journal} {{Rep. Prog. Phys.}}\ }\textbf {\bibinfo {volume}
  {{66}}},\ \bibinfo {pages} {{1937}} (\bibinfo {year} {{2003}})}\BibitemShut
  {NoStop}%
\bibitem [{\citenamefont {Cugliandolo}(2011)}]{Cugliandolo2011}%
  \BibitemOpen
  \bibfield  {author} {\bibinfo {author} {\bibfnamefont {L.~F.}\ \bibnamefont
  {Cugliandolo}},\ }\bibfield  {title} {\bibinfo {title} {The effective
  temperature},\ }\href {https://doi.org/10.1088/1751-8113/44/48/483001}
  {\bibfield  {journal} {\bibinfo  {journal} {J. Phys. A: Math. Theor.}\
  }\textbf {\bibinfo {volume} {44}},\ \bibinfo {pages} {483001} (\bibinfo
  {year} {2011})}\BibitemShut {NoStop}%
\bibitem [{\citenamefont {Takatori}\ and\ \citenamefont
  {Brady}(2015)}]{TakatoriPRE15}%
  \BibitemOpen
  \bibfield  {author} {\bibinfo {author} {\bibfnamefont {S.~C.}\ \bibnamefont
  {Takatori}}\ and\ \bibinfo {author} {\bibfnamefont {J.~F.}\ \bibnamefont
  {Brady}},\ }\bibfield  {title} {\bibinfo {title} {Towards a thermodynamics of
  active matter},\ }\href {https://doi.org/10.1103/PhysRevE.91.032117}
  {\bibfield  {journal} {\bibinfo  {journal} {Phys. Rev. E}\ }\textbf {\bibinfo
  {volume} {91}},\ \bibinfo {pages} {032117} (\bibinfo {year}
  {2015})}\BibitemShut {NoStop}%
\bibitem [{\citenamefont {Solon}\ \emph {et~al.}(2015)\citenamefont {Solon},
  \citenamefont {Fily}, \citenamefont {Baskaran}, \citenamefont {Cates},
  \citenamefont {Kafri}, \citenamefont {Kardar},\ and\ \citenamefont
  {Tailleur}}]{solon2015pressure}%
  \BibitemOpen
  \bibfield  {author} {\bibinfo {author} {\bibfnamefont {A.~P.}\ \bibnamefont
  {Solon}}, \bibinfo {author} {\bibfnamefont {Y.}~\bibnamefont {Fily}},
  \bibinfo {author} {\bibfnamefont {A.}~\bibnamefont {Baskaran}}, \bibinfo
  {author} {\bibfnamefont {M.~E.}\ \bibnamefont {Cates}}, \bibinfo {author}
  {\bibfnamefont {Y.}~\bibnamefont {Kafri}}, \bibinfo {author} {\bibfnamefont
  {M.}~\bibnamefont {Kardar}},\ and\ \bibinfo {author} {\bibfnamefont
  {J.}~\bibnamefont {Tailleur}},\ }\bibfield  {title} {\bibinfo {title}
  {Pressure is not a state function for generic active fluids},\ }\href
  {https://doi.org/10.1038/nphys3377} {\bibfield  {journal} {\bibinfo
  {journal} {Nat. Phys.}\ }\textbf {\bibinfo {volume} {11}},\ \bibinfo {pages}
  {673} (\bibinfo {year} {2015})}\BibitemShut {NoStop}%
\bibitem [{\citenamefont {Paliwal}\ \emph {et~al.}(2018)\citenamefont
  {Paliwal}, \citenamefont {Rodenburg}, \citenamefont {van Roij},\ and\
  \citenamefont {Dijkstra}}]{vanRoij2018}%
  \BibitemOpen
  \bibfield  {author} {\bibinfo {author} {\bibfnamefont {S.}~\bibnamefont
  {Paliwal}}, \bibinfo {author} {\bibfnamefont {J.}~\bibnamefont {Rodenburg}},
  \bibinfo {author} {\bibfnamefont {R.}~\bibnamefont {van Roij}},\ and\
  \bibinfo {author} {\bibfnamefont {M.}~\bibnamefont {Dijkstra}},\ }\bibfield
  {title} {\bibinfo {title} {Chemical potential in active systems: predicting
  phase equilibrium from bulk equations of state?},\ }\href
  {https://doi.org/10.1088/1367-2630/aa9b4d} {\bibfield  {journal} {\bibinfo
  {journal} {New J. Phys.}\ }\textbf {\bibinfo {volume} {20}},\ \bibinfo
  {pages} {015003} (\bibinfo {year} {2018})}\BibitemShut {NoStop}%
\bibitem [{\citenamefont {Speck}(2016)}]{SpeckEL2016}%
  \BibitemOpen
  \bibfield  {author} {\bibinfo {author} {\bibfnamefont {T.}~\bibnamefont
  {Speck}},\ }\bibfield  {title} {\bibinfo {title} {Stochastic thermodynamics
  for active matter},\ }\href {https://doi.org/10.1209/0295-5075/114/30006}
  {\bibfield  {journal} {\bibinfo  {journal} {Europhys. Lett.}\ }\textbf
  {\bibinfo {volume} {114}},\ \bibinfo {pages} {30006} (\bibinfo {year}
  {2016})}\BibitemShut {NoStop}%
\bibitem [{\citenamefont {Einstein}(1905)}]{Einstein1905}%
  \BibitemOpen
  \bibfield  {author} {\bibinfo {author} {\bibfnamefont {A.}~\bibnamefont
  {Einstein}},\ }\bibfield  {title} {\bibinfo {title} {Über die von der
  molekularkinetischen theorie der wärme geforderte bewegung von in ruhenden
  flüssigkeiten suspendierten teilchen},\ }\href
  {https://doi.org/https://doi.org/10.1002/andp.19053220806} {\bibfield
  {journal} {\bibinfo  {journal} {Ann. Phys.}\ }\textbf {\bibinfo {volume}
  {322}},\ \bibinfo {pages} {549} (\bibinfo {year} {1905})}\BibitemShut
  {NoStop}%
\bibitem [{\citenamefont {Kubo}(1966)}]{KuboRPP66}%
  \BibitemOpen
  \bibfield  {author} {\bibinfo {author} {\bibfnamefont {R.}~\bibnamefont
  {Kubo}},\ }\bibfield  {title} {\bibinfo {title} {The fluctuation-dissipation
  theorem},\ }\href {https://doi.org/10.1088/0034-4885/29/1/306} {\bibfield
  {journal} {\bibinfo  {journal} {Rep. Prog. Phys.}\ }\textbf {\bibinfo
  {volume} {29}},\ \bibinfo {pages} {255} (\bibinfo {year} {1966})}\BibitemShut
  {NoStop}%
\bibitem [{\citenamefont {Chen}\ \emph {et~al.}(2006)\citenamefont {Chen},
  \citenamefont {Mallamace}, \citenamefont {Mou}, \citenamefont {Broccio},
  \citenamefont {Corsaro}, \citenamefont {Faraone},\ and\ \citenamefont
  {Liu}}]{ChenPNAS06}%
  \BibitemOpen
  \bibfield  {author} {\bibinfo {author} {\bibfnamefont {S.-H.}\ \bibnamefont
  {Chen}}, \bibinfo {author} {\bibfnamefont {F.}~\bibnamefont {Mallamace}},
  \bibinfo {author} {\bibfnamefont {C.-Y.}\ \bibnamefont {Mou}}, \bibinfo
  {author} {\bibfnamefont {M.}~\bibnamefont {Broccio}}, \bibinfo {author}
  {\bibfnamefont {C.}~\bibnamefont {Corsaro}}, \bibinfo {author} {\bibfnamefont
  {A.}~\bibnamefont {Faraone}},\ and\ \bibinfo {author} {\bibfnamefont
  {L.}~\bibnamefont {Liu}},\ }\bibfield  {title} {\bibinfo {title} {The
  violation of the {S}tokes–{E}instein relation in supercooled water},\
  }\href {https://doi.org/10.1073/pnas.0603253103} {\bibfield  {journal}
  {\bibinfo  {journal} {Proc. Natl. Acade. Sci. U.S.A.}\ }\textbf {\bibinfo
  {volume} {103}},\ \bibinfo {pages} {12974} (\bibinfo {year}
  {2006})}\BibitemShut {NoStop}%
\bibitem [{\citenamefont {Caprini}\ \emph {et~al.}(2024)\citenamefont
  {Caprini}, \citenamefont {Ldov}, \citenamefont {Gupta}, \citenamefont
  {Ellenberg}, \citenamefont {Wittmann}, \citenamefont {L{\"o}wen},\ and\
  \citenamefont {Scholz}}]{CapriniCP24}%
  \BibitemOpen
  \bibfield  {author} {\bibinfo {author} {\bibfnamefont {L.}~\bibnamefont
  {Caprini}}, \bibinfo {author} {\bibfnamefont {A.}~\bibnamefont {Ldov}},
  \bibinfo {author} {\bibfnamefont {R.~K.}\ \bibnamefont {Gupta}}, \bibinfo
  {author} {\bibfnamefont {H.}~\bibnamefont {Ellenberg}}, \bibinfo {author}
  {\bibfnamefont {R.}~\bibnamefont {Wittmann}}, \bibinfo {author}
  {\bibfnamefont {H.}~\bibnamefont {L{\"o}wen}},\ and\ \bibinfo {author}
  {\bibfnamefont {C.}~\bibnamefont {Scholz}},\ }\bibfield  {title} {\bibinfo
  {title} {Emergent memory from tapping collisions in active granular matter},\
  }\href {https://doi.org/10.1038/s42005-024-01540-w} {\bibfield  {journal}
  {\bibinfo  {journal} {Commun. Phys.}\ }\textbf {\bibinfo {volume} {7}},\
  \bibinfo {pages} {52} (\bibinfo {year} {2024})}\BibitemShut {NoStop}%
\bibitem [{\citenamefont {Boriskovsky}\ \emph {et~al.}(2024)\citenamefont
  {Boriskovsky}, \citenamefont {Lindner},\ and\ \citenamefont
  {Roichman}}]{DimaYael24}%
  \BibitemOpen
  \bibfield  {author} {\bibinfo {author} {\bibfnamefont {D.}~\bibnamefont
  {Boriskovsky}}, \bibinfo {author} {\bibfnamefont {B.}~\bibnamefont
  {Lindner}},\ and\ \bibinfo {author} {\bibfnamefont {Y.}~\bibnamefont
  {Roichman}},\ }\href@noop {} {\bibinfo {title} {The fluctuation-dissipation
  relation holds for a macroscopic tracer in an active bath}} (\bibinfo {year}
  {2024}),\ \Eprint {https://arxiv.org/abs/2401.03509} {arXiv:2401.03509}
  \BibitemShut {NoStop}%
\bibitem [{\citenamefont {Geva}\ \emph {et~al.}(2024)\citenamefont {Geva},
  \citenamefont {Admon}, \citenamefont {Levin},\ and\ \citenamefont
  {Roichman}}]{GalorYael24}%
  \BibitemOpen
  \bibfield  {author} {\bibinfo {author} {\bibfnamefont {G.}~\bibnamefont
  {Geva}}, \bibinfo {author} {\bibfnamefont {T.}~\bibnamefont {Admon}},
  \bibinfo {author} {\bibfnamefont {M.}~\bibnamefont {Levin}},\ and\ \bibinfo
  {author} {\bibfnamefont {Y.}~\bibnamefont {Roichman}},\ }\href@noop {}
  {\bibinfo {title} {Diffusive contact between randomly driven colloidal
  suspensions}} (\bibinfo {year} {2024}),\ \Eprint
  {https://arxiv.org/abs/2404.12929} {arXiv:2404.12929} \BibitemShut {NoStop}%
\bibitem [{\citenamefont {del Junco}\ \emph {et~al.}(2018)\citenamefont {del
  Junco}, \citenamefont {Tociu},\ and\ \citenamefont
  {Vaikuntanathan}}]{SuriPNAS18}%
  \BibitemOpen
  \bibfield  {author} {\bibinfo {author} {\bibfnamefont {C.}~\bibnamefont {del
  Junco}}, \bibinfo {author} {\bibfnamefont {T.}~\bibnamefont {Tociu}},\ and\
  \bibinfo {author} {\bibfnamefont {S.}~\bibnamefont {Vaikuntanathan}},\
  }\bibfield  {title} {\bibinfo {title} {Energy dissipation and fluctuations in
  a driven liquid},\ }\href {https://doi.org/10.1073/pnas.1713573115}
  {\bibfield  {journal} {\bibinfo  {journal} {Proc. Natl. Acad. Sci. U.S.A.}\
  }\textbf {\bibinfo {volume} {115}},\ \bibinfo {pages} {3569} (\bibinfo {year}
  {2018})}\BibitemShut {NoStop}%
\bibitem [{\citenamefont {Cugliandolo}\ \emph {et~al.}(1997)\citenamefont
  {Cugliandolo}, \citenamefont {Kurchan},\ and\ \citenamefont
  {Peliti}}]{CuglianoloPRE97}%
  \BibitemOpen
  \bibfield  {author} {\bibinfo {author} {\bibfnamefont {L.~F.}\ \bibnamefont
  {Cugliandolo}}, \bibinfo {author} {\bibfnamefont {J.}~\bibnamefont
  {Kurchan}},\ and\ \bibinfo {author} {\bibfnamefont {L.}~\bibnamefont
  {Peliti}},\ }\bibfield  {title} {\bibinfo {title} {Energy flow, partial
  equilibration, and effective temperatures in systems with slow dynamics},\
  }\href {https://doi.org/10.1103/PhysRevE.55.3898} {\bibfield  {journal}
  {\bibinfo  {journal} {Phys. Rev. E}\ }\textbf {\bibinfo {volume} {55}},\
  \bibinfo {pages} {3898} (\bibinfo {year} {1997})}\BibitemShut {NoStop}%
\bibitem [{\citenamefont {Zamponi}\ \emph {et~al.}()\citenamefont {Zamponi},
  \citenamefont {Bonetto}, \citenamefont {Cugliandolo},\ and\ \citenamefont
  {Kurchan}}]{CugliandoloJSTAT05}%
  \BibitemOpen
  \bibfield  {author} {\bibinfo {author} {\bibfnamefont {F.}~\bibnamefont
  {Zamponi}}, \bibinfo {author} {\bibfnamefont {F.}~\bibnamefont {Bonetto}},
  \bibinfo {author} {\bibfnamefont {L.~F.}\ \bibnamefont {Cugliandolo}},\ and\
  \bibinfo {author} {\bibfnamefont {J.}~\bibnamefont {Kurchan}},\ }\bibfield
  {title} {\bibinfo {title} {A fluctuation theorem for non-equilibrium
  relaxational systems driven by external forces},\ }\href
  {https://doi.org/10.1088/1742-5468/2005/09/P09013} {\bibfield  {journal}
  {\bibinfo  {journal} {J. Stat. Mech.: Theory Exp.}\ }\bibinfo  {number} {
  (2005)},\ \bibinfo {pages} {P09013}}\BibitemShut {NoStop}%
\bibitem [{\citenamefont {Fodor}\ \emph {et~al.}(2016)\citenamefont {Fodor},
  \citenamefont {Nardini}, \citenamefont {Cates}, \citenamefont {Tailleur},
  \citenamefont {Visco},\ and\ \citenamefont {Van~Wijland}}]{fodor2016EP}%
  \BibitemOpen
\bibfield  {number} {  }\bibfield  {author} {\bibinfo {author} {\bibfnamefont
  {{\'E}.}~\bibnamefont {Fodor}}, \bibinfo {author} {\bibfnamefont
  {C.}~\bibnamefont {Nardini}}, \bibinfo {author} {\bibfnamefont {M.~E.}\
  \bibnamefont {Cates}}, \bibinfo {author} {\bibfnamefont {J.}~\bibnamefont
  {Tailleur}}, \bibinfo {author} {\bibfnamefont {P.}~\bibnamefont {Visco}},\
  and\ \bibinfo {author} {\bibfnamefont {F.}~\bibnamefont {Van~Wijland}},\
  }\bibfield  {title} {\bibinfo {title} {How far from equilibrium is active
  matter?},\ }\href {https://doi.org/10.1103/PhysRevLett.117.038103} {\bibfield
   {journal} {\bibinfo  {journal} {Phys. Rev. Lett.}\ }\textbf {\bibinfo
  {volume} {117}},\ \bibinfo {pages} {038103} (\bibinfo {year}
  {2016})}\BibitemShut {NoStop}%
\bibitem [{\citenamefont {Park}\ \emph {et~al.}(2020)\citenamefont {Park},
  \citenamefont {Paneru}, \citenamefont {Kwon}, \citenamefont {Granick},\ and\
  \citenamefont {Pak}}]{GranickSOFT20}%
  \BibitemOpen
  \bibfield  {author} {\bibinfo {author} {\bibfnamefont {J.~T.}\ \bibnamefont
  {Park}}, \bibinfo {author} {\bibfnamefont {G.}~\bibnamefont {Paneru}},
  \bibinfo {author} {\bibfnamefont {C.}~\bibnamefont {Kwon}}, \bibinfo {author}
  {\bibfnamefont {S.}~\bibnamefont {Granick}},\ and\ \bibinfo {author}
  {\bibfnamefont {H.~K.}\ \bibnamefont {Pak}},\ }\bibfield  {title} {\bibinfo
  {title} {Rapid-prototyping a {B}rownian particle in an active bath},\ }\href
  {https://doi.org/10.1039/D0SM00828A} {\bibfield  {journal} {\bibinfo
  {journal} {Soft Matter}\ }\textbf {\bibinfo {volume} {16}},\ \bibinfo {pages}
  {8122} (\bibinfo {year} {2020})}\BibitemShut {NoStop}%
\bibitem [{\citenamefont {Palacci}\ \emph {et~al.}(2010)\citenamefont
  {Palacci}, \citenamefont {Cottin-Bizonne}, \citenamefont {Ybert},\ and\
  \citenamefont {Bocquet}}]{PalacciPRL10}%
  \BibitemOpen
  \bibfield  {author} {\bibinfo {author} {\bibfnamefont {J.}~\bibnamefont
  {Palacci}}, \bibinfo {author} {\bibfnamefont {C.}~\bibnamefont
  {Cottin-Bizonne}}, \bibinfo {author} {\bibfnamefont {C.}~\bibnamefont
  {Ybert}},\ and\ \bibinfo {author} {\bibfnamefont {L.}~\bibnamefont
  {Bocquet}},\ }\bibfield  {title} {\bibinfo {title} {Sedimentation and
  effective temperature of active colloidal suspensions},\ }\href
  {https://doi.org/10.1103/PhysRevLett.105.088304} {\bibfield  {journal}
  {\bibinfo  {journal} {Phys. Rev. Lett.}\ }\textbf {\bibinfo {volume} {105}},\
  \bibinfo {pages} {088304} (\bibinfo {year} {2010})}\BibitemShut {NoStop}%
\bibitem [{\citenamefont {Granek}\ \emph {et~al.}(2022)\citenamefont {Granek},
  \citenamefont {Kafri},\ and\ \citenamefont {Tailleur}}]{GranekPRL22}%
  \BibitemOpen
  \bibfield  {author} {\bibinfo {author} {\bibfnamefont {O.}~\bibnamefont
  {Granek}}, \bibinfo {author} {\bibfnamefont {Y.}~\bibnamefont {Kafri}},\ and\
  \bibinfo {author} {\bibfnamefont {J.}~\bibnamefont {Tailleur}},\ }\bibfield
  {title} {\bibinfo {title} {Anomalous transport of tracers in active baths},\
  }\href {https://doi.org/10.1103/PhysRevLett.129.038001} {\bibfield  {journal}
  {\bibinfo  {journal} {Phys. Rev. Lett.}\ }\textbf {\bibinfo {volume} {129}},\
  \bibinfo {pages} {038001} (\bibinfo {year} {2022})}\BibitemShut {NoStop}%
\bibitem [{\citenamefont {Solon}\ and\ \citenamefont
  {Horowitz}(2022)}]{SolonJPA22}%
  \BibitemOpen
  \bibfield  {author} {\bibinfo {author} {\bibfnamefont {A.}~\bibnamefont
  {Solon}}\ and\ \bibinfo {author} {\bibfnamefont {J.~M.}\ \bibnamefont
  {Horowitz}},\ }\bibfield  {title} {\bibinfo {title} {On the {E}instein
  relation between mobility and diffusion coefficient in an active bath},\
  }\href {https://doi.org/10.1088/1751-8121/ac5d82} {\bibfield  {journal}
  {\bibinfo  {journal} {J. Phys. A: Math. Theor.}\ }\textbf {\bibinfo {volume}
  {55}},\ \bibinfo {pages} {184002} (\bibinfo {year} {2022})}\BibitemShut
  {NoStop}%
\bibitem [{\citenamefont {Basu}\ \emph {et~al.}(2008)\citenamefont {Basu},
  \citenamefont {Joanny}, \citenamefont {J\"ulicher},\ and\ \citenamefont
  {Prost}}]{BasuEPJE08}%
  \BibitemOpen
  \bibfield  {author} {\bibinfo {author} {\bibfnamefont {A.}~\bibnamefont
  {Basu}}, \bibinfo {author} {\bibfnamefont {J.~F.}\ \bibnamefont {Joanny}},
  \bibinfo {author} {\bibfnamefont {F.}~\bibnamefont {J\"ulicher}},\ and\
  \bibinfo {author} {\bibfnamefont {J.}~\bibnamefont {Prost}},\ }\bibfield
  {title} {\bibinfo {title} {Thermal and non-thermal fluctuations in active
  polar gels},\ }\bibfield  {journal} {\bibinfo  {journal} {Eur. Phys. J. E}\
  }\textbf {\bibinfo {volume} {27}},\ \href
  {https://doi.org/10.1140/epje/i2008-10364-9} {10.1140/epje/i2008-10364-9}
  (\bibinfo {year} {2008})\BibitemShut {NoStop}%
\bibitem [{\citenamefont {Mizuno}\ \emph {et~al.}(2007)\citenamefont {Mizuno},
  \citenamefont {Tardin}, \citenamefont {Schmidt},\ and\ \citenamefont
  {MacKintosh}}]{FredSCI07}%
  \BibitemOpen
  \bibfield  {author} {\bibinfo {author} {\bibfnamefont {D.}~\bibnamefont
  {Mizuno}}, \bibinfo {author} {\bibfnamefont {C.}~\bibnamefont {Tardin}},
  \bibinfo {author} {\bibfnamefont {C.~F.}\ \bibnamefont {Schmidt}},\ and\
  \bibinfo {author} {\bibfnamefont {F.~C.}\ \bibnamefont {MacKintosh}},\
  }\bibfield  {title} {\bibinfo {title} {Nonequilibrium mechanics of active
  cytoskeletal networks},\ }\href {https://doi.org/10.1126/science.1134404}
  {\bibfield  {journal} {\bibinfo  {journal} {Science}\ }\textbf {\bibinfo
  {volume} {315}},\ \bibinfo {pages} {370} (\bibinfo {year}
  {2007})}\BibitemShut {NoStop}%
\bibitem [{\citenamefont {Chen}\ \emph {et~al.}(2020)\citenamefont {Chen},
  \citenamefont {Markovich},\ and\ \citenamefont {MacKintosh}}]{Chen2020}%
  \BibitemOpen
  \bibfield  {author} {\bibinfo {author} {\bibfnamefont {S.}~\bibnamefont
  {Chen}}, \bibinfo {author} {\bibfnamefont {T.}~\bibnamefont {Markovich}},\
  and\ \bibinfo {author} {\bibfnamefont {F.~C.}\ \bibnamefont {MacKintosh}},\
  }\bibfield  {title} {\bibinfo {title} {Motor-free contractility in active
  gels},\ }\href {https://doi.org/10.1103/PhysRevLett.125.208101} {\bibfield
  {journal} {\bibinfo  {journal} {Phys. Rev. Lett.}\ }\textbf {\bibinfo
  {volume} {125}},\ \bibinfo {pages} {208101} (\bibinfo {year}
  {2020})}\BibitemShut {NoStop}%
\bibitem [{\citenamefont {Chen}\ \emph {et~al.}(2007)\citenamefont {Chen},
  \citenamefont {Lau}, \citenamefont {Hough}, \citenamefont {Islam},
  \citenamefont {Goulian}, \citenamefont {Lubensky},\ and\ \citenamefont
  {Yodh}}]{ChenPRL07}%
  \BibitemOpen
  \bibfield  {author} {\bibinfo {author} {\bibfnamefont {D.~T.~N.}\
  \bibnamefont {Chen}}, \bibinfo {author} {\bibfnamefont {A.~W.~C.}\
  \bibnamefont {Lau}}, \bibinfo {author} {\bibfnamefont {L.~A.}\ \bibnamefont
  {Hough}}, \bibinfo {author} {\bibfnamefont {M.~F.}\ \bibnamefont {Islam}},
  \bibinfo {author} {\bibfnamefont {M.}~\bibnamefont {Goulian}}, \bibinfo
  {author} {\bibfnamefont {T.~C.}\ \bibnamefont {Lubensky}},\ and\ \bibinfo
  {author} {\bibfnamefont {A.~G.}\ \bibnamefont {Yodh}},\ }\bibfield  {title}
  {\bibinfo {title} {Fluctuations and rheology in active bacterial
  suspensions},\ }\href {https://doi.org/10.1103/PhysRevLett.99.148302}
  {\bibfield  {journal} {\bibinfo  {journal} {Phys. Rev. Lett.}\ }\textbf
  {\bibinfo {volume} {99}},\ \bibinfo {pages} {148302} (\bibinfo {year}
  {2007})}\BibitemShut {NoStop}%
\bibitem [{\citenamefont {Maggi}\ \emph {et~al.}(2014)\citenamefont {Maggi},
  \citenamefont {Paoluzzi}, \citenamefont {Pellicciotta}, \citenamefont
  {Lepore}, \citenamefont {Angelani},\ and\ \citenamefont
  {Di~Leonardo}}]{MaggiPRL14}%
  \BibitemOpen
  \bibfield  {author} {\bibinfo {author} {\bibfnamefont {C.}~\bibnamefont
  {Maggi}}, \bibinfo {author} {\bibfnamefont {M.}~\bibnamefont {Paoluzzi}},
  \bibinfo {author} {\bibfnamefont {N.}~\bibnamefont {Pellicciotta}}, \bibinfo
  {author} {\bibfnamefont {A.}~\bibnamefont {Lepore}}, \bibinfo {author}
  {\bibfnamefont {L.}~\bibnamefont {Angelani}},\ and\ \bibinfo {author}
  {\bibfnamefont {R.}~\bibnamefont {Di~Leonardo}},\ }\bibfield  {title}
  {\bibinfo {title} {Generalized energy equipartition in harmonic oscillators
  driven by active baths},\ }\href
  {https://doi.org/10.1103/PhysRevLett.113.238303} {\bibfield  {journal}
  {\bibinfo  {journal} {Phys. Rev. Lett.}\ }\textbf {\bibinfo {volume} {113}},\
  \bibinfo {pages} {238303} (\bibinfo {year} {2014})}\BibitemShut {NoStop}%
\bibitem [{\citenamefont {Lau}\ \emph {et~al.}(2007)\citenamefont {Lau},
  \citenamefont {Lacoste},\ and\ \citenamefont {Mallick}}]{LacostePRL07}%
  \BibitemOpen
  \bibfield  {author} {\bibinfo {author} {\bibfnamefont {A.~W.~C.}\
  \bibnamefont {Lau}}, \bibinfo {author} {\bibfnamefont {D.}~\bibnamefont
  {Lacoste}},\ and\ \bibinfo {author} {\bibfnamefont {K.}~\bibnamefont
  {Mallick}},\ }\bibfield  {title} {\bibinfo {title} {Nonequilibrium
  fluctuations and mechanochemical couplings of a molecular motor},\ }\href
  {https://doi.org/10.1103/PhysRevLett.99.158102} {\bibfield  {journal}
  {\bibinfo  {journal} {Phys. Rev. Lett.}\ }\textbf {\bibinfo {volume} {99}},\
  \bibinfo {pages} {158102} (\bibinfo {year} {2007})}\BibitemShut {NoStop}%
\bibitem [{\citenamefont {Peliti}\ and\ \citenamefont
  {Pigolotti}(2021)}]{book:stochastic}%
  \BibitemOpen
  \bibfield  {author} {\bibinfo {author} {\bibfnamefont {L.}~\bibnamefont
  {Peliti}}\ and\ \bibinfo {author} {\bibfnamefont {S.}~\bibnamefont
  {Pigolotti}},\ }\href@noop {} {\emph {\bibinfo {title} {Stochastic
  Thermodynamics}}}\ (\bibinfo  {publisher} {Princeton University Press},\
  \bibinfo {year} {2021})\BibitemShut {NoStop}%
\bibitem [{\citenamefont {Sekimoto}(1998)}]{sekimoto98Langevin_thermo}%
  \BibitemOpen
  \bibfield  {author} {\bibinfo {author} {\bibfnamefont {K.}~\bibnamefont
  {Sekimoto}},\ }\bibfield  {title} {\bibinfo {title} {{Langevin Equation and
  Thermodynamics}},\ }\href {https://doi.org/10.1143/PTPS.130.17} {\bibfield
  {journal} {\bibinfo  {journal} {Prog. Theor. Phys.}\ }\textbf {\bibinfo
  {volume} {130}},\ \bibinfo {pages} {17} (\bibinfo {year} {1998})}\BibitemShut
  {NoStop}%
\bibitem [{\citenamefont {Sekimoto}(2010)}]{book:sekimoto}%
  \BibitemOpen
  \bibfield  {author} {\bibinfo {author} {\bibfnamefont {K.}~\bibnamefont
  {Sekimoto}},\ }\href@noop {} {\emph {\bibinfo {title} {Stochastic
  Energetics}}},\ Lecture Notes in Physics\ (\bibinfo  {publisher} {Springer},\
  \bibinfo {year} {2010})\BibitemShut {NoStop}%
\bibitem [{\citenamefont {Markovich}\ \emph {et~al.}(2021)\citenamefont
  {Markovich}, \citenamefont {Fodor}, \citenamefont {Tjhung},\ and\
  \citenamefont {Cates}}]{PRX2021}%
  \BibitemOpen
  \bibfield  {author} {\bibinfo {author} {\bibfnamefont {T.}~\bibnamefont
  {Markovich}}, \bibinfo {author} {\bibfnamefont {E.}~\bibnamefont {Fodor}},
  \bibinfo {author} {\bibfnamefont {E.}~\bibnamefont {Tjhung}},\ and\ \bibinfo
  {author} {\bibfnamefont {M.~E.}\ \bibnamefont {Cates}},\ }\bibfield  {title}
  {\bibinfo {title} {Thermodynamics of active field theories: Energetic cost of
  coupling to reservoirs},\ }\href {https://doi.org/10.1103/PhysRevX.11.021057}
  {\bibfield  {journal} {\bibinfo  {journal} {Phys. Rev. X}\ }\textbf {\bibinfo
  {volume} {11}},\ \bibinfo {pages} {021057} (\bibinfo {year}
  {2021})}\BibitemShut {NoStop}%
\bibitem [{\citenamefont {Seifert}(2012)}]{SeifertRPP12}%
  \BibitemOpen
  \bibfield  {author} {\bibinfo {author} {\bibfnamefont {U.}~\bibnamefont
  {Seifert}},\ }\bibfield  {title} {\bibinfo {title} {Stochastic
  thermodynamics, fluctuation theorems and molecular machines},\ }\href
  {https://doi.org/10.1088/0034-4885/75/12/126001} {\bibfield  {journal}
  {\bibinfo  {journal} {Rep. Prog. Phys.}\ }\textbf {\bibinfo {volume} {75}},\
  \bibinfo {pages} {126001} (\bibinfo {year} {2012})}\BibitemShut {NoStop}%
\bibitem [{\citenamefont {Jarzynski}(1997{\natexlab{a}})}]{JarzynskiPRL97}%
  \BibitemOpen
  \bibfield  {author} {\bibinfo {author} {\bibfnamefont {C.}~\bibnamefont
  {Jarzynski}},\ }\bibfield  {title} {\bibinfo {title} {Nonequilibrium equality
  for free energy differences},\ }\href
  {https://doi.org/10.1103/PhysRevLett.78.2690} {\bibfield  {journal} {\bibinfo
   {journal} {Phys. Rev. Lett.}\ }\textbf {\bibinfo {volume} {78}},\ \bibinfo
  {pages} {2690} (\bibinfo {year} {1997}{\natexlab{a}})}\BibitemShut {NoStop}%
\bibitem [{\citenamefont {Jarzynski}(1997{\natexlab{b}})}]{JarzynskiPRE97}%
  \BibitemOpen
  \bibfield  {author} {\bibinfo {author} {\bibfnamefont {C.}~\bibnamefont
  {Jarzynski}},\ }\bibfield  {title} {\bibinfo {title} {Equilibrium free-energy
  differences from nonequilibrium measurements: A master-equation approach},\
  }\href {https://doi.org/10.1103/PhysRevE.56.5018} {\bibfield  {journal}
  {\bibinfo  {journal} {Phys. Rev. E}\ }\textbf {\bibinfo {volume} {56}},\
  \bibinfo {pages} {5018} (\bibinfo {year} {1997}{\natexlab{b}})}\BibitemShut
  {NoStop}%
\bibitem [{\citenamefont {Crooks}(1999)}]{CrooksPRE99}%
  \BibitemOpen
  \bibfield  {author} {\bibinfo {author} {\bibfnamefont {G.~E.}\ \bibnamefont
  {Crooks}},\ }\bibfield  {title} {\bibinfo {title} {Entropy production
  fluctuation theorem and the nonequilibrium work relation for free energy
  differences},\ }\href {https://doi.org/10.1103/PhysRevE.60.2721} {\bibfield
  {journal} {\bibinfo  {journal} {Phys. Rev. E}\ }\textbf {\bibinfo {volume}
  {60}},\ \bibinfo {pages} {2721} (\bibinfo {year} {1999})}\BibitemShut
  {NoStop}%
\bibitem [{\citenamefont {Nardini}\ \emph {et~al.}(2017)\citenamefont
  {Nardini}, \citenamefont {Fodor}, \citenamefont {Tjhung}, \citenamefont {van
  Wijland}, \citenamefont {Tailleur},\ and\ \citenamefont
  {Cates}}]{NardiniPRX17}%
  \BibitemOpen
  \bibfield  {author} {\bibinfo {author} {\bibfnamefont {C.}~\bibnamefont
  {Nardini}}, \bibinfo {author} {\bibfnamefont {E.}~\bibnamefont {Fodor}},
  \bibinfo {author} {\bibfnamefont {E.}~\bibnamefont {Tjhung}}, \bibinfo
  {author} {\bibfnamefont {F.}~\bibnamefont {van Wijland}}, \bibinfo {author}
  {\bibfnamefont {J.}~\bibnamefont {Tailleur}},\ and\ \bibinfo {author}
  {\bibfnamefont {M.~E.}\ \bibnamefont {Cates}},\ }\bibfield  {title} {\bibinfo
  {title} {Entropy production in field theories without time-reversal symmetry:
  Quantifying the non-equilibrium character of active matter},\ }\href
  {https://doi.org/10.1103/PhysRevX.7.021007} {\bibfield  {journal} {\bibinfo
  {journal} {Phys. Rev. X}\ }\textbf {\bibinfo {volume} {7}},\ \bibinfo {pages}
  {021007} (\bibinfo {year} {2017})}\BibitemShut {NoStop}%
\bibitem [{\citenamefont {Martin}\ \emph {et~al.}(2021)\citenamefont {Martin},
  \citenamefont {O'Byrne}, \citenamefont {Cates}, \citenamefont {Fodor},
  \citenamefont {Nardini}, \citenamefont {Tailleur},\ and\ \citenamefont {van
  Wijland}}]{MartinPRE21}%
  \BibitemOpen
  \bibfield  {author} {\bibinfo {author} {\bibfnamefont {D.}~\bibnamefont
  {Martin}}, \bibinfo {author} {\bibfnamefont {J.}~\bibnamefont {O'Byrne}},
  \bibinfo {author} {\bibfnamefont {M.~E.}\ \bibnamefont {Cates}}, \bibinfo
  {author} {\bibfnamefont {E.}~\bibnamefont {Fodor}}, \bibinfo {author}
  {\bibfnamefont {C.}~\bibnamefont {Nardini}}, \bibinfo {author} {\bibfnamefont
  {J.}~\bibnamefont {Tailleur}},\ and\ \bibinfo {author} {\bibfnamefont
  {F.}~\bibnamefont {van Wijland}},\ }\bibfield  {title} {\bibinfo {title}
  {Statistical mechanics of active {O}rnstein-{U}hlenbeck particles},\ }\href
  {https://doi.org/10.1103/PhysRevE.103.032607} {\bibfield  {journal} {\bibinfo
   {journal} {Phys. Rev. E}\ }\textbf {\bibinfo {volume} {103}},\ \bibinfo
  {pages} {032607} (\bibinfo {year} {2021})}\BibitemShut {NoStop}%
\bibitem [{\citenamefont {Khadem}\ \emph {et~al.}(2022)\citenamefont {Khadem},
  \citenamefont {Klages},\ and\ \citenamefont {Klapp}}]{RainerPRR22}%
  \BibitemOpen
  \bibfield  {author} {\bibinfo {author} {\bibfnamefont {S.~M.~J.}\
  \bibnamefont {Khadem}}, \bibinfo {author} {\bibfnamefont {R.}~\bibnamefont
  {Klages}},\ and\ \bibinfo {author} {\bibfnamefont {S.~H.~L.}\ \bibnamefont
  {Klapp}},\ }\bibfield  {title} {\bibinfo {title} {Stochastic thermodynamics
  of fractional brownian motion},\ }\href
  {https://doi.org/10.1103/PhysRevResearch.4.043186} {\bibfield  {journal}
  {\bibinfo  {journal} {Phys. Rev. Res.}\ }\textbf {\bibinfo {volume} {4}},\
  \bibinfo {pages} {043186} (\bibinfo {year} {2022})}\BibitemShut {NoStop}%
\bibitem [{\citenamefont {Chechkin}\ \emph {et~al.}()\citenamefont {Chechkin},
  \citenamefont {Lenz},\ and\ \citenamefont {Klages}}]{RainerJSTAT12}%
  \BibitemOpen
  \bibfield  {author} {\bibinfo {author} {\bibfnamefont {A.~V.}\ \bibnamefont
  {Chechkin}}, \bibinfo {author} {\bibfnamefont {F.}~\bibnamefont {Lenz}},\
  and\ \bibinfo {author} {\bibfnamefont {R.}~\bibnamefont {Klages}},\
  }\bibfield  {title} {\bibinfo {title} {Normal and anomalous fluctuation
  relations for {G}aussian stochastic dynamics},\ }\href
  {https://doi.org/10.1088/1742-5468/2012/11/L11001} {\bibfield  {journal}
  {\bibinfo  {journal} {J. Stat. Mech.: Theory Exp.}\ }\bibinfo  {number} {
  (2012)},\ \bibinfo {pages} {L11001}}\BibitemShut {NoStop}%
\bibitem [{\citenamefont {Hatano}\ and\ \citenamefont
  {Sasa}(2001)}]{HatanoPRL01}%
  \BibitemOpen
\bibfield  {number} {  }\bibfield  {author} {\bibinfo {author} {\bibfnamefont
  {T.}~\bibnamefont {Hatano}}\ and\ \bibinfo {author} {\bibfnamefont
  {S.}~\bibnamefont {Sasa}},\ }\bibfield  {title} {\bibinfo {title}
  {Steady-state thermodynamics of langevin systems},\ }\href
  {https://doi.org/10.1103/PhysRevLett.86.3463} {\bibfield  {journal} {\bibinfo
   {journal} {Phys. Rev. Lett.}\ }\textbf {\bibinfo {volume} {86}},\ \bibinfo
  {pages} {3463} (\bibinfo {year} {2001})}\BibitemShut {NoStop}%
\bibitem [{\citenamefont {Trepagnier}\ \emph {et~al.}(2004)\citenamefont
  {Trepagnier}, \citenamefont {Jarzynski}, \citenamefont {Ritort},
  \citenamefont {Crooks}, \citenamefont {Bustamante},\ and\ \citenamefont
  {Liphardt}}]{RitortPNAS04}%
  \BibitemOpen
  \bibfield  {author} {\bibinfo {author} {\bibfnamefont {E.~H.}\ \bibnamefont
  {Trepagnier}}, \bibinfo {author} {\bibfnamefont {C.}~\bibnamefont
  {Jarzynski}}, \bibinfo {author} {\bibfnamefont {F.}~\bibnamefont {Ritort}},
  \bibinfo {author} {\bibfnamefont {G.~E.}\ \bibnamefont {Crooks}}, \bibinfo
  {author} {\bibfnamefont {C.~J.}\ \bibnamefont {Bustamante}},\ and\ \bibinfo
  {author} {\bibfnamefont {J.}~\bibnamefont {Liphardt}},\ }\bibfield  {title}
  {\bibinfo {title} {Experimental test of hatano and sasa's nonequilibrium
  steady-state equality},\ }\href {https://doi.org/10.1073/pnas.0406405101}
  {\bibfield  {journal} {\bibinfo  {journal} {Proc. Natl. Acad. Sci. U.S.A.}\
  }\textbf {\bibinfo {volume} {101}},\ \bibinfo {pages} {15038} (\bibinfo
  {year} {2004})}\BibitemShut {NoStop}%
\bibitem [{Dis()}]{DissipForce}%
  \BibitemOpen
  \href@noop {} {}\bibinfo {note} {Nonconservative forces may depend on
  velocities as well. Here, we consider the overdamped limit where velocity
  dependencies average out. The position-dependent forces lead indeed to heat
  dissipation, as seen in Eq.~\eqref{eq:dQ}.}\BibitemShut {Stop}%
\bibitem [{\citenamefont {Schuss}(2010)}]{book:schuss}%
  \BibitemOpen
  \bibfield  {author} {\bibinfo {author} {\bibfnamefont {Z.}~\bibnamefont
  {Schuss}},\ }\href@noop {} {\emph {\bibinfo {title} {Theory and Applications
  of Stochastic Processes}}}\ (\bibinfo  {publisher} {Springer},\ \bibinfo
  {address} {New York},\ \bibinfo {year} {2010})\BibitemShut {NoStop}%
\bibitem [{\citenamefont {Lau}\ and\ \citenamefont
  {Lubensky}(2007)}]{LubenskyPRE07}%
  \BibitemOpen
  \bibfield  {author} {\bibinfo {author} {\bibfnamefont {A.~W.~C.}\
  \bibnamefont {Lau}}\ and\ \bibinfo {author} {\bibfnamefont {T.~C.}\
  \bibnamefont {Lubensky}},\ }\bibfield  {title} {\bibinfo {title}
  {State-dependent diffusion: Thermodynamic consistency and its path integral
  formulation},\ }\href {https://doi.org/10.1103/PhysRevE.76.011123} {\bibfield
   {journal} {\bibinfo  {journal} {Phys. Rev. E}\ }\textbf {\bibinfo {volume}
  {76}},\ \bibinfo {pages} {011123} (\bibinfo {year} {2007})}\BibitemShut
  {NoStop}%
\bibitem [{\citenamefont {Cates}\ \emph {et~al.}(2022)\citenamefont {Cates},
  \citenamefont {Fodor}, \citenamefont {Markovich}, \citenamefont {Nardini},\
  and\ \citenamefont {Tjhung}}]{CatesEnt22}%
  \BibitemOpen
  \bibfield  {author} {\bibinfo {author} {\bibfnamefont {M.~E.}\ \bibnamefont
  {Cates}}, \bibinfo {author} {\bibfnamefont {E.}~\bibnamefont {Fodor}},
  \bibinfo {author} {\bibfnamefont {T.}~\bibnamefont {Markovich}}, \bibinfo
  {author} {\bibfnamefont {C.}~\bibnamefont {Nardini}},\ and\ \bibinfo {author}
  {\bibfnamefont {E.}~\bibnamefont {Tjhung}},\ }\bibfield  {title} {\bibinfo
  {title} {Stochastic hydrodynamics of complex fluids: Discretisation and
  entropy production},\ }\href {https://doi.org/10.3390/e24020254} {\bibfield
  {journal} {\bibinfo  {journal} {Entropy}\ }\textbf {\bibinfo {volume} {24}},\
  \bibinfo {pages} {254} (\bibinfo {year} {2022})}\BibitemShut {NoStop}%
\bibitem [{\citenamefont {Risken}(1996)}]{book:FPE}%
  \BibitemOpen
  \bibfield  {author} {\bibinfo {author} {\bibfnamefont {H.}~\bibnamefont
  {Risken}},\ }\href@noop {} {\emph {\bibinfo {title} {The Fokker-Planck
  Equation}}}\ (\bibinfo  {publisher} {Springer Berlin},\ \bibinfo {address}
  {Heidelberg},\ \bibinfo {year} {1996})\BibitemShut {NoStop}%
\bibitem [{STR()}]{STRATO}%
  \BibitemOpen
  \href@noop {} {}\bibinfo {note} {From a physical perspective, if the forces
  are conservative (derived from a Hamiltonian), then the integral over forces
  must give the change in energy. This means that the ``usual'' chain rule must
  hold. Mathematically, this implies using the {S}tratonovich product
  rule~\cite{book:schuss,book:stochastic}. For nonconservative forces, they
  arise from coarse graining of fast degrees of freedom for which the forces
  were conservative~\cite{JarzynskiPRL97}.}\BibitemShut {Stop}%
\bibitem [{NON()}]{NONEQENT}%
  \BibitemOpen
  \href@noop {} {}\bibinfo {note} {Although it appears to be consistent within
  many stochastic thermodynamic theories~\cite{book:stochastic}, since entorpy
  is originally an equilibrium property, there is a debate regarding the
  adequacy of the Shannon entropy as a microscopic functional for the
  thermodynamic entorpy in out-of-equilibrium
  scenarios~\cite{LebowitzPA99,GoldsteinPD04,LevinRP14}.}\BibitemShut {Stop}%
\bibitem [{\citenamefont {Shannon}(1948)}]{Shannon1948}%
  \BibitemOpen
  \bibfield  {author} {\bibinfo {author} {\bibfnamefont {C.~E.}\ \bibnamefont
  {Shannon}},\ }\bibfield  {title} {\bibinfo {title} {{A mathematical theory of
  communication}},\ }\href
  {https://doi.org/{10.1002/j.1538-7305.1948.tb01338.x}} {\bibfield  {journal}
  {\bibinfo  {journal} {{Bell Syst. Tech. J.}}\ }\textbf {\bibinfo {volume}
  {{27}}},\ \bibinfo {pages} {379} (\bibinfo {year} {{1948}})}\BibitemShut
  {NoStop}%
\bibitem [{\citenamefont {Cover}\ and\ \citenamefont
  {Thomas}(2006)}]{BOOK:infotheor}%
  \BibitemOpen
  \bibfield  {author} {\bibinfo {author} {\bibfnamefont {T.~M.}\ \bibnamefont
  {Cover}}\ and\ \bibinfo {author} {\bibfnamefont {J.~A.}\ \bibnamefont
  {Thomas}},\ }\href@noop {} {\emph {\bibinfo {title} {Elements of information
  theory}}},\ \bibinfo {edition} {2nd}\ ed.\ (\bibinfo  {publisher}
  {Wiley-Interscience},\ \bibinfo {address} {Hoboken, New Jersey},\ \bibinfo
  {year} {2006})\BibitemShut {NoStop}%
\bibitem [{\citenamefont {Kardar}(2007)}]{book:kardar1}%
  \BibitemOpen
  \bibfield  {author} {\bibinfo {author} {\bibfnamefont {M.}~\bibnamefont
  {Kardar}},\ }\href@noop {} {\emph {\bibinfo {title} {Statistical Physics of
  Particles}}}\ (\bibinfo  {publisher} {Cambridge University Press},\ \bibinfo
  {year} {2007})\BibitemShut {NoStop}%
\bibitem [{\citenamefont {Seifert}(2005)}]{SeifertPRL05}%
  \BibitemOpen
  \bibfield  {author} {\bibinfo {author} {\bibfnamefont {U.}~\bibnamefont
  {Seifert}},\ }\bibfield  {title} {\bibinfo {title} {Entropy production along
  a stochastic trajectory and an integral fluctuation theorem},\ }\href
  {https://doi.org/10.1103/PhysRevLett.95.040602} {\bibfield  {journal}
  {\bibinfo  {journal} {Phys. Rev. Lett.}\ }\textbf {\bibinfo {volume} {95}},\
  \bibinfo {pages} {040602} (\bibinfo {year} {2005})}\BibitemShut {NoStop}%
\bibitem [{\citenamefont {Jarzynski}(1999)}]{JarzynskiJSP99}%
  \BibitemOpen
  \bibfield  {author} {\bibinfo {author} {\bibfnamefont {C.}~\bibnamefont
  {Jarzynski}},\ }\bibfield  {title} {\bibinfo {title} {Microscopic analysis of
  clausius–duhem processes},\ }\href
  {https://doi.org/10.1023/A:1004541004050} {\bibfield  {journal} {\bibinfo
  {journal} {J. Stat. Phys.}\ }\textbf {\bibinfo {volume} {96}},\ \bibinfo
  {pages} {415} (\bibinfo {year} {1999})}\BibitemShut {NoStop}%
\bibitem [{\citenamefont {Mart\'inez}\ \emph {et~al.}(2019)\citenamefont
  {Mart\'inez}, \citenamefont {Bisker}, \citenamefont {Horowitz},\ and\
  \citenamefont {Parrondo}}]{BiskerNATCOM19}%
  \BibitemOpen
  \bibfield  {author} {\bibinfo {author} {\bibfnamefont {I.~A.}\ \bibnamefont
  {Mart\'inez}}, \bibinfo {author} {\bibfnamefont {G.}~\bibnamefont {Bisker}},
  \bibinfo {author} {\bibfnamefont {J.~M.}\ \bibnamefont {Horowitz}},\ and\
  \bibinfo {author} {\bibfnamefont {J.~M.~R.}\ \bibnamefont {Parrondo}},\
  }\bibfield  {title} {\bibinfo {title} {Inferring broken detailed balance in
  the absence of observable currents},\ }\href
  {https://doi.org/10.1038/s41467-019-11051-w} {\bibfield  {journal} {\bibinfo
  {journal} {Nat. Commun.}\ }\textbf {\bibinfo {volume} {10}},\ \bibinfo
  {pages} {3542} (\bibinfo {year} {2019})}\BibitemShut {NoStop}%
\bibitem [{\citenamefont {Skinner}\ and\ \citenamefont
  {Dunkel}(2021)}]{dunkelPRL21}%
  \BibitemOpen
  \bibfield  {author} {\bibinfo {author} {\bibfnamefont {D.~J.}\ \bibnamefont
  {Skinner}}\ and\ \bibinfo {author} {\bibfnamefont {J.}~\bibnamefont
  {Dunkel}},\ }\bibfield  {title} {\bibinfo {title} {Estimating entropy
  production from waiting time distributions},\ }\href
  {https://doi.org/10.1103/PhysRevLett.127.198101} {\bibfield  {journal}
  {\bibinfo  {journal} {Phys. Rev. Lett.}\ }\textbf {\bibinfo {volume} {127}},\
  \bibinfo {pages} {198101} (\bibinfo {year} {2021})}\BibitemShut {NoStop}%
\bibitem [{\citenamefont {Sorkin}\ \emph {et~al.}(2024)\citenamefont {Sorkin},
  \citenamefont {Ariel},\ and\ \citenamefont {Markovich}}]{JSTAT24}%
  \BibitemOpen
  \bibfield  {author} {\bibinfo {author} {\bibfnamefont {B.}~\bibnamefont
  {Sorkin}}, \bibinfo {author} {\bibfnamefont {G.}~\bibnamefont {Ariel}},\ and\
  \bibinfo {author} {\bibfnamefont {T.}~\bibnamefont {Markovich}},\ }\href@noop
  {} {\bibinfo {title} {Consistent expansion of the langevin propagator with
  application to entropy production}} (\bibinfo {year} {2024}),\ \Eprint
  {https://arxiv.org/abs/2405.13855} {arXiv:2405.13855} \BibitemShut {NoStop}%
\bibitem [{Ref()}]{RefCatesFoot}%
  \BibitemOpen
  \href@noop {} {}\bibinfo {note} {This expression can also be obtained from
  Ref.~\cite[Eq.~(30)]{CatesEnt22} by replacing their $\Gamma\cal U'$ with $\mu
  F$.}\BibitemShut {Stop}%
\bibitem [{\citenamefont {Onsager}(1931{\natexlab{a}})}]{onsagerPR31b}%
  \BibitemOpen
  \bibfield  {author} {\bibinfo {author} {\bibfnamefont {L.}~\bibnamefont
  {Onsager}},\ }\bibfield  {title} {\bibinfo {title} {Reciprocal relations in
  irreversible processes. ii.},\ }\href
  {https://doi.org/10.1103/PhysRev.38.2265} {\bibfield  {journal} {\bibinfo
  {journal} {Phys. Rev.}\ }\textbf {\bibinfo {volume} {38}},\ \bibinfo {pages}
  {2265} (\bibinfo {year} {1931}{\natexlab{a}})}\BibitemShut {NoStop}%
\bibitem [{\citenamefont {Onsager}(1931{\natexlab{b}})}]{onsagerPR31a}%
  \BibitemOpen
  \bibfield  {author} {\bibinfo {author} {\bibfnamefont {L.}~\bibnamefont
  {Onsager}},\ }\bibfield  {title} {\bibinfo {title} {Reciprocal relations in
  irreversible processes. i.},\ }\href {https://doi.org/10.1103/PhysRev.37.405}
  {\bibfield  {journal} {\bibinfo  {journal} {Phys. Rev.}\ }\textbf {\bibinfo
  {volume} {37}},\ \bibinfo {pages} {405} (\bibinfo {year}
  {1931}{\natexlab{b}})}\BibitemShut {NoStop}%
\bibitem [{\citenamefont {\"Ottinger}(2005)}]{BOOK:Ottinger}%
  \BibitemOpen
  \bibfield  {author} {\bibinfo {author} {\bibfnamefont {H.~S.}\ \bibnamefont
  {\"Ottinger}},\ }\href@noop {} {\emph {\bibinfo {title} {Beyond Equilibrium
  Thermodynamics}}}\ (\bibinfo  {publisher} {Wiley},\ \bibinfo {year}
  {2005})\BibitemShut {NoStop}%
\bibitem [{Sei()}]{SeifertFOOT}%
  \BibitemOpen
  \href@noop {} {}\bibinfo {note} {This expression has already appeared in
  {R}ef.~\cite{SeifertPRL05}, albeit in a simpler scenario.}\BibitemShut
  {Stop}%
\bibitem [{\citenamefont {Pigolotti}\ \emph {et~al.}(2017)\citenamefont
  {Pigolotti}, \citenamefont {Neri}, \citenamefont {Rold\'an},\ and\
  \citenamefont {J\"ulicher}}]{PigolottiPRL17}%
  \BibitemOpen
  \bibfield  {author} {\bibinfo {author} {\bibfnamefont {S.}~\bibnamefont
  {Pigolotti}}, \bibinfo {author} {\bibfnamefont {I.}~\bibnamefont {Neri}},
  \bibinfo {author} {\bibfnamefont {E.}~\bibnamefont {Rold\'an}},\ and\
  \bibinfo {author} {\bibfnamefont {F.}~\bibnamefont {J\"ulicher}},\ }\bibfield
   {title} {\bibinfo {title} {Generic properties of stochastic entropy
  production},\ }\href {https://doi.org/10.1103/PhysRevLett.119.140604}
  {\bibfield  {journal} {\bibinfo  {journal} {Phys. Rev. Lett.}\ }\textbf
  {\bibinfo {volume} {119}},\ \bibinfo {pages} {140604} (\bibinfo {year}
  {2017})}\BibitemShut {NoStop}%
\bibitem [{\citenamefont {Harada}\ and\ \citenamefont
  {Sasa}(2005)}]{HaradaPRL05}%
  \BibitemOpen
  \bibfield  {author} {\bibinfo {author} {\bibfnamefont {T.}~\bibnamefont
  {Harada}}\ and\ \bibinfo {author} {\bibfnamefont {S.}~\bibnamefont {Sasa}},\
  }\bibfield  {title} {\bibinfo {title} {Equality connecting energy dissipation
  with a violation of the fluctuation-response relation},\ }\href
  {https://doi.org/10.1103/PhysRevLett.95.130602} {\bibfield  {journal}
  {\bibinfo  {journal} {Phys. Rev. Lett.}\ }\textbf {\bibinfo {volume} {95}},\
  \bibinfo {pages} {130602} (\bibinfo {year} {2005})}\BibitemShut {NoStop}%
\bibitem [{\citenamefont {Baiesi}\ \emph {et~al.}(2009)\citenamefont {Baiesi},
  \citenamefont {Maes},\ and\ \citenamefont {Wynants}}]{BaiesiJSP2009}%
  \BibitemOpen
  \bibfield  {author} {\bibinfo {author} {\bibfnamefont {M.}~\bibnamefont
  {Baiesi}}, \bibinfo {author} {\bibfnamefont {C.}~\bibnamefont {Maes}},\ and\
  \bibinfo {author} {\bibfnamefont {B.}~\bibnamefont {Wynants}},\ }\bibfield
  {title} {\bibinfo {title} {Nonequilibrium linear response for markov
  dynamics, {I}: Jump processes and overdamped diffusions},\ }\href
  {https://doi.org/10.1007/s10955-009-9852-8} {\bibfield  {journal} {\bibinfo
  {journal} {J. Stat. Phys.}\ }\textbf {\bibinfo {volume} {137}},\ \bibinfo
  {pages} {1094} (\bibinfo {year} {2009})}\BibitemShut {NoStop}%
\bibitem [{SM()}]{SM}%
  \BibitemOpen
  \href@noop {} {}\bibinfo {note} {See {S}upplemental {M}aterial for the
  derivation and violation of the Fluctuation-dissipation theorem and
  Harada-Sasa relation~\cite{HaradaPRL05}, further discussion on classical
  thermodynamic theorems, and further details on nonequilibrium steady-states
  with multiple temparatures. There we also cite
  Refs.~\cite{book:KloedenPlaten,book:zwanzig,FDTderiv,GolestanianARXIV2024,AgarwalZph72}.}\BibitemShut
  {Stop}%
\bibitem [{\citenamefont {Datta}\ \emph {et~al.}(2022)\citenamefont {Datta},
  \citenamefont {Pietzonka},\ and\ \citenamefont {Barato}}]{DattaPRX22}%
  \BibitemOpen
  \bibfield  {author} {\bibinfo {author} {\bibfnamefont {A.}~\bibnamefont
  {Datta}}, \bibinfo {author} {\bibfnamefont {P.}~\bibnamefont {Pietzonka}},\
  and\ \bibinfo {author} {\bibfnamefont {A.~C.}\ \bibnamefont {Barato}},\
  }\bibfield  {title} {\bibinfo {title} {Second law for active heat engines},\
  }\href {https://doi.org/10.1103/PhysRevX.12.031034} {\bibfield  {journal}
  {\bibinfo  {journal} {Phys. Rev. X}\ }\textbf {\bibinfo {volume} {12}},\
  \bibinfo {pages} {031034} (\bibinfo {year} {2022})}\BibitemShut {NoStop}%
\bibitem [{\citenamefont {Grosberg}\ and\ \citenamefont
  {Joanny}(2015)}]{JoannyPRE15}%
  \BibitemOpen
  \bibfield  {author} {\bibinfo {author} {\bibfnamefont {A.~Y.}\ \bibnamefont
  {Grosberg}}\ and\ \bibinfo {author} {\bibfnamefont {J.-F.}\ \bibnamefont
  {Joanny}},\ }\bibfield  {title} {\bibinfo {title} {Nonequilibrium statistical
  mechanics of mixtures of particles in contact with different thermostats},\
  }\href {https://doi.org/10.1103/PhysRevE.92.032118} {\bibfield  {journal}
  {\bibinfo  {journal} {Phys. Rev. E}\ }\textbf {\bibinfo {volume} {92}},\
  \bibinfo {pages} {032118} (\bibinfo {year} {2015})}\BibitemShut {NoStop}%
\bibitem [{\citenamefont {Weber}\ \emph {et~al.}(2016)\citenamefont {Weber},
  \citenamefont {Weber},\ and\ \citenamefont {Frey}}]{FreyPRL16}%
  \BibitemOpen
  \bibfield  {author} {\bibinfo {author} {\bibfnamefont {S.~N.}\ \bibnamefont
  {Weber}}, \bibinfo {author} {\bibfnamefont {C.~A.}\ \bibnamefont {Weber}},\
  and\ \bibinfo {author} {\bibfnamefont {E.}~\bibnamefont {Frey}},\ }\bibfield
  {title} {\bibinfo {title} {Binary mixtures of particles with different
  diffusivities demix},\ }\href
  {https://doi.org/10.1103/PhysRevLett.116.058301} {\bibfield  {journal}
  {\bibinfo  {journal} {Phys. Rev. Lett.}\ }\textbf {\bibinfo {volume} {116}},\
  \bibinfo {pages} {058301} (\bibinfo {year} {2016})}\BibitemShut {NoStop}%
\bibitem [{\citenamefont {Stenhammar}\ \emph {et~al.}(2015)\citenamefont
  {Stenhammar}, \citenamefont {Wittkowski}, \citenamefont {Marenduzzo},\ and\
  \citenamefont {Cates}}]{CatesPRL15}%
  \BibitemOpen
  \bibfield  {author} {\bibinfo {author} {\bibfnamefont {J.}~\bibnamefont
  {Stenhammar}}, \bibinfo {author} {\bibfnamefont {R.}~\bibnamefont
  {Wittkowski}}, \bibinfo {author} {\bibfnamefont {D.}~\bibnamefont
  {Marenduzzo}},\ and\ \bibinfo {author} {\bibfnamefont {M.~E.}\ \bibnamefont
  {Cates}},\ }\bibfield  {title} {\bibinfo {title} {Activity-induced phase
  separation and self-assembly in mixtures of active and passive particles},\
  }\href {https://doi.org/10.1103/PhysRevLett.114.018301} {\bibfield  {journal}
  {\bibinfo  {journal} {Phys. Rev. Lett.}\ }\textbf {\bibinfo {volume} {114}},\
  \bibinfo {pages} {018301} (\bibinfo {year} {2015})}\BibitemShut {NoStop}%
\bibitem [{\citenamefont {Benisty}\ \emph {et~al.}(2015)\citenamefont
  {Benisty}, \citenamefont {Ben-Jacob}, \citenamefont {Ariel},\ and\
  \citenamefont {Be'er}}]{ArielPRL15}%
  \BibitemOpen
  \bibfield  {author} {\bibinfo {author} {\bibfnamefont {S.}~\bibnamefont
  {Benisty}}, \bibinfo {author} {\bibfnamefont {E.}~\bibnamefont {Ben-Jacob}},
  \bibinfo {author} {\bibfnamefont {G.}~\bibnamefont {Ariel}},\ and\ \bibinfo
  {author} {\bibfnamefont {A.}~\bibnamefont {Be'er}},\ }\bibfield  {title}
  {\bibinfo {title} {Antibiotic-induced anomalous statistics of collective
  bacterial swarming},\ }\href {https://doi.org/10.1103/PhysRevLett.114.018105}
  {\bibfield  {journal} {\bibinfo  {journal} {Phys. Rev. Lett.}\ }\textbf
  {\bibinfo {volume} {114}},\ \bibinfo {pages} {018105} (\bibinfo {year}
  {2015})}\BibitemShut {NoStop}%
\bibitem [{\citenamefont {Brahmachari}\ \emph {et~al.}(2024)\citenamefont
  {Brahmachari}, \citenamefont {Markovich}, \citenamefont {MacKintosh},\ and\
  \citenamefont {Onuchic}}]{MarkovichBIO23}%
  \BibitemOpen
  \bibfield  {author} {\bibinfo {author} {\bibfnamefont {S.}~\bibnamefont
  {Brahmachari}}, \bibinfo {author} {\bibfnamefont {T.}~\bibnamefont
  {Markovich}}, \bibinfo {author} {\bibfnamefont {F.~C.}\ \bibnamefont
  {MacKintosh}},\ and\ \bibinfo {author} {\bibfnamefont {J.~N.}\ \bibnamefont
  {Onuchic}},\ }\bibfield  {title} {\bibinfo {title} {Temporally correlated
  active forces drive segregation and enhanced dynamics in chromosome
  polymers},\ }\href {https://doi.org/10.1103/PRXLife.2.033003} {\bibfield
  {journal} {\bibinfo  {journal} {PRX Life}\ }\textbf {\bibinfo {volume} {2}},\
  \bibinfo {pages} {033003} (\bibinfo {year} {2024})}\BibitemShut {NoStop}%
\bibitem [{\citenamefont {Ariel}\ \emph {et~al.}(2015)\citenamefont {Ariel},
  \citenamefont {Rimer},\ and\ \citenamefont {Ben-Jacob}}]{ArielJSP15}%
  \BibitemOpen
  \bibfield  {author} {\bibinfo {author} {\bibfnamefont {G.}~\bibnamefont
  {Ariel}}, \bibinfo {author} {\bibfnamefont {O.}~\bibnamefont {Rimer}},\ and\
  \bibinfo {author} {\bibfnamefont {E.}~\bibnamefont {Ben-Jacob}},\ }\bibfield
  {title} {\bibinfo {title} {Order–disorder phase transition in heterogeneous
  populations of self-propelled particles},\ }\href
  {https://doi.org/10.1007/s10955-014-1095-7} {\bibfield  {journal} {\bibinfo
  {journal} {J. Stat. Phys.}\ }\textbf {\bibinfo {volume} {158}},\ \bibinfo
  {pages} {579} (\bibinfo {year} {2015})}\BibitemShut {NoStop}%
\bibitem [{\citenamefont {Netzer}\ \emph {et~al.}(2019)\citenamefont {Netzer},
  \citenamefont {Yarom},\ and\ \citenamefont {Ariel}}]{ArielPHYSA19}%
  \BibitemOpen
  \bibfield  {author} {\bibinfo {author} {\bibfnamefont {G.}~\bibnamefont
  {Netzer}}, \bibinfo {author} {\bibfnamefont {Y.}~\bibnamefont {Yarom}},\ and\
  \bibinfo {author} {\bibfnamefont {G.}~\bibnamefont {Ariel}},\ }\bibfield
  {title} {\bibinfo {title} {Heterogeneous populations in a network model of
  collective motion},\ }\href {https://doi.org/10.1016/j.physa.2019.121550}
  {\bibfield  {journal} {\bibinfo  {journal} {Physica A: Stat. Mech. Appl.}\
  }\textbf {\bibinfo {volume} {530}},\ \bibinfo {pages} {121550} (\bibinfo
  {year} {2019})}\BibitemShut {NoStop}%
\bibitem [{\citenamefont {Bisker}\ \emph {et~al.}()\citenamefont {Bisker},
  \citenamefont {Polettini}, \citenamefont {Gingrich},\ and\ \citenamefont
  {Horowitz}}]{BiskerJSM2017}%
  \BibitemOpen
  \bibfield  {author} {\bibinfo {author} {\bibfnamefont {G.}~\bibnamefont
  {Bisker}}, \bibinfo {author} {\bibfnamefont {M.}~\bibnamefont {Polettini}},
  \bibinfo {author} {\bibfnamefont {T.~R.}\ \bibnamefont {Gingrich}},\ and\
  \bibinfo {author} {\bibfnamefont {J.~M.}\ \bibnamefont {Horowitz}},\
  }\bibfield  {title} {\bibinfo {title} {Hierarchical bounds on entropy
  production inferred from partial information},\ }\href
  {https://doi.org/10.1088/1742-5468/aa8c0d} {\bibfield  {journal} {\bibinfo
  {journal} {J. Stat. Mech.: Theory Exp.}\ }\bibinfo  {number} { (2017)},\
  \bibinfo {pages} {093210}}\BibitemShut {NoStop}%
\bibitem [{\citenamefont {Kawai}\ \emph {et~al.}(2007)\citenamefont {Kawai},
  \citenamefont {Parrondo},\ and\ \citenamefont {den Broeck}}]{Parrondo07EPR}%
  \BibitemOpen
\bibfield  {number} {  }\bibfield  {author} {\bibinfo {author} {\bibfnamefont
  {R.}~\bibnamefont {Kawai}}, \bibinfo {author} {\bibfnamefont {J.~M.~R.}\
  \bibnamefont {Parrondo}},\ and\ \bibinfo {author} {\bibfnamefont {C.~V.}\
  \bibnamefont {den Broeck}},\ }\bibfield  {title} {\bibinfo {title}
  {Dissipation: The phase-space perspective},\ }\href
  {https://doi.org/10.1103/PhysRevLett.98.080602} {\bibfield  {journal}
  {\bibinfo  {journal} {Phys. Rev. Lett.}\ }\textbf {\bibinfo {volume} {98}},\
  \bibinfo {pages} {080602} (\bibinfo {year} {2007})}\BibitemShut {NoStop}%
\bibitem [{\citenamefont {Gingrich}\ \emph {et~al.}(2016)\citenamefont
  {Gingrich}, \citenamefont {Horowitz}, \citenamefont {Perunov},\ and\
  \citenamefont {England}}]{GingrichPRL16}%
  \BibitemOpen
  \bibfield  {author} {\bibinfo {author} {\bibfnamefont {T.~R.}\ \bibnamefont
  {Gingrich}}, \bibinfo {author} {\bibfnamefont {J.~M.}\ \bibnamefont
  {Horowitz}}, \bibinfo {author} {\bibfnamefont {N.}~\bibnamefont {Perunov}},\
  and\ \bibinfo {author} {\bibfnamefont {J.~L.}\ \bibnamefont {England}},\
  }\bibfield  {title} {\bibinfo {title} {Dissipation bounds all steady-state
  current fluctuations},\ }\href
  {https://doi.org/10.1103/PhysRevLett.116.120601} {\bibfield  {journal}
  {\bibinfo  {journal} {Phys. Rev. Lett.}\ }\textbf {\bibinfo {volume} {116}},\
  \bibinfo {pages} {120601} (\bibinfo {year} {2016})}\BibitemShut {NoStop}%
\bibitem [{\citenamefont {Barato}\ and\ \citenamefont
  {Seifert}(2015)}]{SeifertPRL15}%
  \BibitemOpen
  \bibfield  {author} {\bibinfo {author} {\bibfnamefont {A.~C.}\ \bibnamefont
  {Barato}}\ and\ \bibinfo {author} {\bibfnamefont {U.}~\bibnamefont
  {Seifert}},\ }\bibfield  {title} {\bibinfo {title} {Thermodynamic uncertainty
  relation for biomolecular processes},\ }\href
  {https://doi.org/10.1103/PhysRevLett.114.158101} {\bibfield  {journal}
  {\bibinfo  {journal} {Phys. Rev. Lett.}\ }\textbf {\bibinfo {volume} {114}},\
  \bibinfo {pages} {158101} (\bibinfo {year} {2015})}\BibitemShut {NoStop}%
\bibitem [{\citenamefont {Lebowitz}(1999)}]{LebowitzPA99}%
  \BibitemOpen
  \bibfield  {author} {\bibinfo {author} {\bibfnamefont {J.~L.}\ \bibnamefont
  {Lebowitz}},\ }\bibfield  {title} {\bibinfo {title} {Microscopic origins of
  irreversible macroscopic behavior},\ }\href
  {https://doi.org/https://doi.org/10.1016/S0378-4371(98)00514-7} {\bibfield
  {journal} {\bibinfo  {journal} {Phys. A: Stat. Mech.}\ }\textbf {\bibinfo
  {volume} {263}},\ \bibinfo {pages} {516} (\bibinfo {year}
  {1999})}\BibitemShut {NoStop}%
\bibitem [{\citenamefont {Goldstein}\ and\ \citenamefont
  {Lebowitz}(2004)}]{GoldsteinPD04}%
  \BibitemOpen
  \bibfield  {author} {\bibinfo {author} {\bibfnamefont {S.}~\bibnamefont
  {Goldstein}}\ and\ \bibinfo {author} {\bibfnamefont {J.~L.}\ \bibnamefont
  {Lebowitz}},\ }\bibfield  {title} {\bibinfo {title} {On the ({B}oltzmann)
  entropy of non-equilibrium systems},\ }\href
  {https://doi.org/https://doi.org/10.1016/j.physd.2004.01.008} {\bibfield
  {journal} {\bibinfo  {journal} {Physica D}\ }\textbf {\bibinfo {volume}
  {193}},\ \bibinfo {pages} {53} (\bibinfo {year} {2004})}\BibitemShut
  {NoStop}%
\bibitem [{\citenamefont {Levin}\ \emph {et~al.}(2014)\citenamefont {Levin},
  \citenamefont {Pakter}, \citenamefont {Rizzato}, \citenamefont {Teles},\ and\
  \citenamefont {Benetti}}]{LevinRP14}%
  \BibitemOpen
  \bibfield  {author} {\bibinfo {author} {\bibfnamefont {Y.}~\bibnamefont
  {Levin}}, \bibinfo {author} {\bibfnamefont {P.}~\bibnamefont {Pakter}},
  \bibinfo {author} {\bibfnamefont {F.~B.}\ \bibnamefont {Rizzato}}, \bibinfo
  {author} {\bibfnamefont {T.~N.}\ \bibnamefont {Teles}},\ and\ \bibinfo
  {author} {\bibfnamefont {F.~P.~C.}\ \bibnamefont {Benetti}},\ }\bibfield
  {title} {\bibinfo {title} {Nonequilibrium statistical mechanics of systems
  with long-range interactions},\ }\href
  {https://doi.org/https://doi.org/10.1016/j.physrep.2013.10.001} {\bibfield
  {journal} {\bibinfo  {journal} {Phys. Rep.}\ }\textbf {\bibinfo {volume}
  {535}},\ \bibinfo {pages} {1} (\bibinfo {year} {2014})}\BibitemShut {NoStop}%
\bibitem [{\citenamefont {Kloeden}\ and\ \citenamefont
  {Platen}(1992)}]{book:KloedenPlaten}%
  \BibitemOpen
  \bibfield  {author} {\bibinfo {author} {\bibfnamefont {P.~E.}\ \bibnamefont
  {Kloeden}}\ and\ \bibinfo {author} {\bibfnamefont {E.}~\bibnamefont
  {Platen}},\ }\href@noop {} {\emph {\bibinfo {title} {Numerical Solution of
  Stochastic Differential Equations}}},\ Stochastic Modelling and Applied
  Probability\ (\bibinfo  {publisher} {Springer Berlin},\ \bibinfo {address}
  {Heidelberg},\ \bibinfo {year} {1992})\BibitemShut {NoStop}%
\bibitem [{\citenamefont {Zwanzig}(2001)}]{book:zwanzig}%
  \BibitemOpen
  \bibfield  {author} {\bibinfo {author} {\bibfnamefont {R.}~\bibnamefont
  {Zwanzig}},\ }\href@noop {} {\emph {\bibinfo {title} {Nonequilibrium
  Statistical Mechanics}}}\ (\bibinfo  {publisher} {Oxford University Press},\
  \bibinfo {address} {New York},\ \bibinfo {year} {2001})\BibitemShut {NoStop}%
\bibitem [{\citenamefont {Baiesi}\ and\ \citenamefont {Maes}(2013)}]{FDTderiv}%
  \BibitemOpen
  \bibfield  {author} {\bibinfo {author} {\bibfnamefont {M.}~\bibnamefont
  {Baiesi}}\ and\ \bibinfo {author} {\bibfnamefont {C.}~\bibnamefont {Maes}},\
  }\bibfield  {title} {\bibinfo {title} {An update on the nonequilibrium linear
  response},\ }\href {https://doi.org/10.1088/1367-2630/15/1/013004} {\bibfield
   {journal} {\bibinfo  {journal} {New J. Phys.}\ }\textbf {\bibinfo {volume}
  {15}},\ \bibinfo {pages} {013004} (\bibinfo {year} {2013})}\BibitemShut
  {NoStop}%
\bibitem [{\citenamefont {Golestanian}(2024)}]{GolestanianARXIV2024}%
  \BibitemOpen
  \bibfield  {author} {\bibinfo {author} {\bibfnamefont {R.}~\bibnamefont
  {Golestanian}},\ }\href@noop {} {\bibinfo {title} {Hydrodynamically
  consistent many-body {H}arada-{S}asa relation}} (\bibinfo {year} {2024}),\
  \Eprint {https://arxiv.org/abs/2406.16761} {arXiv:2406.16761} \BibitemShut
  {NoStop}%
\bibitem [{\citenamefont {Agarwal}(1972)}]{AgarwalZph72}%
  \BibitemOpen
  \bibfield  {author} {\bibinfo {author} {\bibfnamefont {G.~S.}\ \bibnamefont
  {Agarwal}},\ }\bibfield  {title} {\bibinfo {title} {Fluctuation-dissipation
  theorems for systems in non-thermal equilibrium and applications},\ }\href
  {https://doi.org/10.1007/BF01391621} {\bibfield  {journal} {\bibinfo
  {journal} {Z. Phys.}\ }\textbf {\bibinfo {volume} {252}},\ \bibinfo {pages}
  {25} (\bibinfo {year} {1972})}\BibitemShut {NoStop}%
\end{thebibliography}
\end{document}